\documentclass[aip,amsmath,amssymb,reprint]{revtex4-1}
\usepackage{graphicx}
\usepackage{dcolumn}
\usepackage{bm}
\usepackage{ulem}
\usepackage[utf8]{inputenc}
\usepackage[T1]{fontenc}
\usepackage{mathptmx}
\usepackage{etoolbox}
\usepackage{xcolor}
\usepackage{multirow}

\makeatletter
\def\@email#1#2{%
 \endgroup
 \patchcmd{\titleblock@produce}
  {\frontmatter@RRAPformat}
  {\frontmatter@RRAPformat{\produce@RRAP{*#1\href{mailto:#2}{#2}}}\frontmatter@RRAPformat}
  {}{}
}%
\makeatother
\begin{document}

\preprint{}

\title[Correlations versus noise in the NFT market]{Correlations versus noise in the NFT market}

\author{Marcin W{\k a}torek}
 \email{marcin.watorek@pk.edu.pl}
\author{Pawe{\l} Szyd{\l}o}

\affiliation{Faculty of Computer Science and Telecommunications, Cracow University of Technology, ul.~Warszawska 24, 31-155 Krak\'ow, Poland}

\author{Jaros{\l}aw Kwapie\'n}

\affiliation{Complex Systems Theory Department, Institute of Nuclear Physics, Polish Academy of Sciences, ul.~Radzikowskiego 152, 31-342 Krak\'ow, Poland}

\author{Stanis{\l}aw Dro\.zd\.z}

\affiliation{Faculty of Computer Science and Telecommunications, Cracow University of Technology, ul.~Warszawska 24, 31-155 Krak\'ow, Poland}
\affiliation{Complex Systems Theory Department, Institute of Nuclear Physics, Polish Academy of Sciences, ul.~Radzikowskiego 152, 31-342 Krak\'ow, Poland}

\date{\today}

\begin{abstract}
The non-fungible token (NFT) market emerges as a recent trading innovation leveraging blockchain technology, mirroring the dynamics of the cryptocurrency market. The current study is based on the capitalization changes and transaction volumes across a large number of token collections on the Ethereum platform. In order to deepen the understanding of the market dynamics, the collection-collection dependencies are examined by using the multivariate formalism of detrended correlation coefficient and correlation matrix. It appears that correlation strength is lower here than that observed in previously studied markets. Consequently, the eigenvalue spectra of the correlation matrix more closely follow the Marchenko-Pastur distribution, still, some departures indicating the existence of correlations remain. The comparison of results obtained from the correlation matrix built from the Pearson coefficients and, independently, from the detrended cross-correlation coefficients suggests that the global correlations in the NFT market arise from higher frequency fluctuations. Corresponding minimal spanning trees (MSTs) for capitalization variability exhibit a scale-free character while, for the number of transactions, they are somewhat more decentralized.
\end{abstract}

\maketitle

\begin{quotation}
The complexity of financial markets, encompassing traditional assets and emerging forms like cryptocurrencies and non-fungible tokens (NFTs), is a multi-scale phenomenon driven by various factors. Understanding this complexity and the associated emergent phenomena requires delving into the details and intricacies of these markets, dynamic interaction of diverse elements, and the evolving system of global finance. NFTs represent a unique form of digital assets, which may be tied to digital art, collectibles, or virtual real estate. The additional complexity of the NFT markets arise from the innovative aspects of digital asset valuation, intellectual property considerations, and the other aspects of smart contracts. NFTs also introduce new possibilities for creators and collectors, but come with challenges related to copyright, authenticity, and market saturation. The issue of global correlations within NFT markets has not been quantitatively addressed in the literature so far and therefore, in this study, it is framed in the context of the potential competition between correlation effects and randomness.
\end{quotation}

\section{Introduction}

The cryptoassets space is constantly evolving. Since the first cryptocurrency - Bitcoin - launched in 2009~\cite{NakamotoS-2009a}, blockchain technology has undergone considerable evolution and found applications in various fields~\cite{blockchainapp}. The initial functionality of Bitcoin and its early alternatives was limited to settling transactions and serving as a distributed ledger register. This changed with the launch of Ethereum in 2015 as a decentralized, open-source, and distributed computing platform~\cite{Ethereum}. This marked the first significant expansion of blockchain capabilities. The Ethereum smart contracts functionality allowed for the easy creation of one's own fungible tokens in ERC-20 standard~\cite{ERC-20}. Among the early applications of smart contracts was one of the first and highest-valued NFT collections - CryptoPunks. However, the concept of unique tokens is older and can be traced back to Bitcoin colored coins in 2012~\cite{Rosenfeld2013OverviewOC} and the creation of the first-ever NFT, Quantum, minted by Kevin McCoy on Namecoin in 2014~\cite{quantmNFT}. The Ethereum project experienced significant growth and development, particularly in 2017, when numerous projects launched on its blockchain through initial coin offerings (ICOs), ultimately contributing to a market bubble~\cite{BELLON2022}. At the peak of the ICO bubble in December 2017, the challenges in the capacity of the Ethereum network became increasingly evident. The surge in popularity of the CryptoKitties game exacerbated these issues, at times consuming up to 70\% of Ethereum's usage capacity. This spike in demand led to record highs in the number of transactions and fees, considerably slowing down the network~\cite{Jiang2021}. The game enables participants to purchase, sell, and create non-fungible tokens (NFTs) on the Ethereum blockchain. It stands out as one of the earliest and most prominent examples of the ERC-721 standard, which became officially recognized as the Non-Fungible Token standard in 2018~\cite{ERC-721}. This standard facilitates the straightforward creation of NFT tokens and collections by anyone.

The application of smart contracts to finance marked a significant second milestone for the cryptoasset domain, giving rise to the phenomenon known as DeFi (Decentralized Finance) Summer that lasted between mid-2020 and early 2021~\cite{Maouchi2022}. This was accompanied by the development of blockchains that competed with Ethereum, such as Solana, Avalanche, Arbitrum, BNB Chain, Polygon, Tron, Optimism, and Cardano. The total value locked in DeFi protocols on various blockchains peaked in Dec 2021 with a value of 250 billion USD~\cite{TVL}. In this period, NFT tokens surged in popularity and entered the mainstream, driven by a low barrier to entry, the simplicity of launching personal collections, and celebrity endorsements. These factors fueled the speculative NFT bubble of 2021~\cite{Guo2023}. The boom started in March 2021 when Mike Winkelmann, aka Beeple, sold his NFT Everydays: The First 5,000 Days (2021) for \$69 million at Christie’s~\cite{Christies}. Shortly after that, the 10,000-part series Bored Ape Yacht Club was launched in May 2021, starting a wave of similar projects called profile-pic NFTs (PFP NFTs). Many celebrities owned a Bored Ape Yacht Club NFT or released their own collections~\cite{HORKY2023,kim2024,NFTceleb} and even the former US president Donald Trump had participated in the NFT collective enthusiasm~\cite{TrumpNFT}. After spectacular growth, when the most expensive NFT token (CryptoPunk $\#5822$) was sold for 8000 ETH in February 2022,  the NFT market eventually peaked in May 2022, reaching a total capitalization of $\sim$21 billion USD~\cite{NFTbubble}. The widespread popularity of NFTs at that time allowed for funding support for Ukraine via selling one of the CryptoPunk tokens in June 2022 for 90 ETH ($\sim$100,000 USD). After the subsequent crash~\cite{NFTcrash}, the total capitalization dropped to 5 billion USD in April 2024~\cite{coinmarketcap}. Given the unregulated nature of the NFT market, it has been subject to numerous manipulative practices~\cite{bose2023exploiting,Sifat2024} including wash trading~\cite{vonWachterV-2022a} and pump-and-dump schemes~\cite{bose2023}.

Despite a dramatic drop of nearly 90\% in market value, the blockchain technology underpinning NFTs continues to evolve. The Bitcoin Ordinals protocol was introduced to the Bitcoin network in January 2023~\cite{BTCordinals}, enabling the creation of collections on the Bitcoin blockchain. In December 2023, this blockchain accounted for the largest portion of NFT trading volume at 42.1\%. Nonetheless, throughout 2023, Ethereum maintained its position as the dominant blockchain for NFT transactions with 72.3\% trading volume share~\cite{Blokchianshare}. Other blockchains hosting NFT collections include Solana, Immutable X, Polygon, BNB Chain, Flow, Arbitrum, Avalanche, and Ronin. There are many NFT marketplaces that enable NFT collections trading; the three largest ones are Blur, Magic Eaden and OpenSea as of April 2024~\cite{NFTmarketplaces}. Although the current main NFT application to pure speculation faced serious criticism~\cite{NFTcrit} and, despite many challenges~\cite{Ali2023}, the standard may still bring a new outlook in how digital assets are perceived, owned, and transacted, reshaping the digital world's understanding of value and ownership~\cite{Chalmers2022,giannoni2023blockchain}.

Building on our previous exploration of price fluctuation characteristics within individual NFT collections~\cite{Szydlo2024}, this article aims to delve into the correlations among NFT collection trades. The well-established financial markets, such as stocks~\cite{James2021chaos,James2022PhysAstock}, Forex~\cite{drozdz2007world,gkebarowski2019detecting,Miskiewicz2021}, and even younger cryptocurrencies~\cite{STOSIC2018,Chaudhari2020,Chaos2020,watorek2021,James2022,Nguyen2022,gavin2023community,James2023}, demonstrate significant collective behavior in their cross-correlations. Random Matrix Theory (RMT) has been effectively utilized to uncover the nature and origins of correlations in these markets~\cite{Laloux1999,Plerou1999,Drozdz2001PhysA}. In this study, RMT is applied to examine the price and the number of transactions variations of the collections created on the Ethereum blockchain. This analysis will focus on determining the non-random nature of these correlations and examining if distinct NFT-market sectors can be distinguished. Properties of minimal spanning trees constructed from the correlation networks will also be analyzed. This approach has already been successfully applied to financial time series~\cite{MANTEGNA1999,tumminello2010correlation}, including stocks~\cite{MICCICHE2003,WILINSKI2013,Miskiewicz2022}, currencies~\cite{gorski2008scale,gkebarowski2019detecting}, and cryptocurrencies~\cite{POLOVNIKOV2020,watorek2021,WatorekM-2023b,brigatti2023inferring,NGUYEN2023}.

\section{Data and methods}

\subsection{Data specification}
\label{Dataspec}
The dataset contains tick-by-tick data representing transactions (price and time) on 90 NFT collections from the Ethereum blockchain. It covers the period of $T=500$ days that began on June 9, 2022 and ended on October 21, 2023. The data was sourced from the CryptoSlam! portal~\cite{CryptoSlam}. The collections were selected according to the criterion requiring that, on average, each collection had at least two transactions involving at least one of its tokens. Each dataset was then converted into a bivariate time series representing (1) the collection capitalization logarithmic increments $c_{\Delta t}(t) = \ln K(t+\Delta t)-\ln K(t)$ expressed in USD, where $K(t)$ is the total capitalization at a moment $t$, and (2) the number $N_{\Delta t}$ of transactions in time unit $\Delta t$. Two sampling frequencies were chosen: $\Delta t = 1$h, suitable for the 13 most liquid collections where less than 50\% of hours had no trading activity (see the column $\%0_{\Delta t=1{\rm h}}$ in Tab.~\ref{NFTnames}), and $\Delta t = 24$h, applicable to all collections. Transaction prices in various cryptocurrencies were also available for extraction. However, since transactions on the Ethereum blockchain occur not only in ETH but also in a multitude of other tokens, this information proved to be impractical for consolidation into a single time series. The complete list of the collections, along with their basic characteristics such as capitalization on the dataset's final day, the total number of transactions, collection size measured by the number of circulated tokens, and a fraction of the hours without transactions, can be found in Tab.~\ref{NFTnames}. 

\begin{figure}
\includegraphics[width=0.49\textwidth]{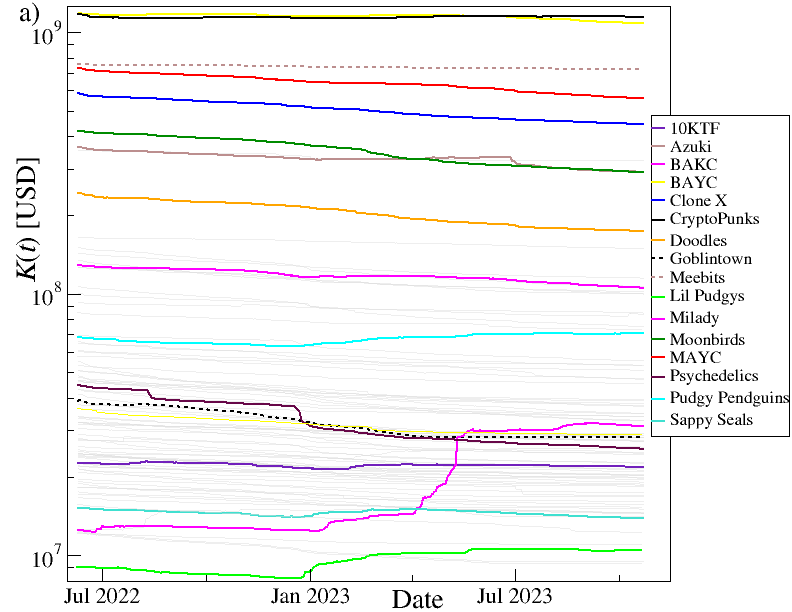}
\includegraphics[width=0.49\textwidth]{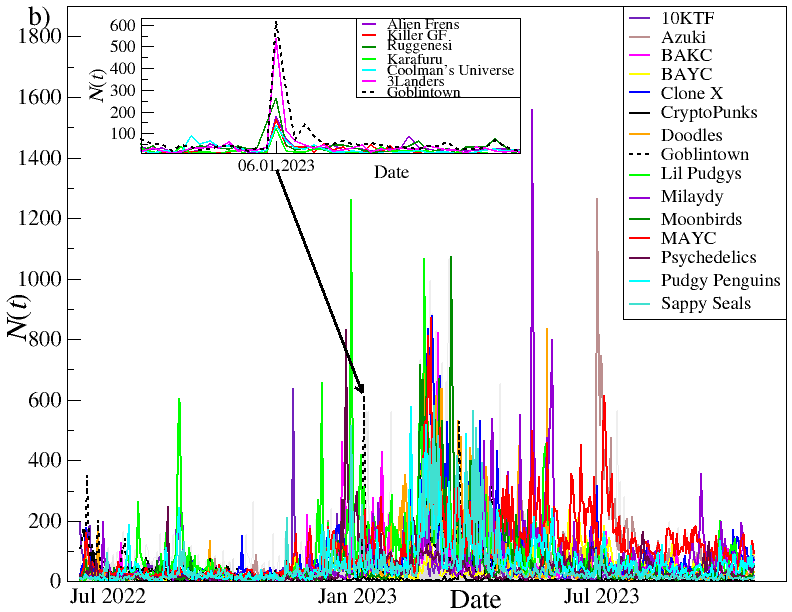}
\caption{(a) Collection capitalization $K(t)$ and (b) the number of transactions $N_{\Delta t=24h}(t)$ for all the considered NFT collections with the most frequently traded ones listed explicitly. Inset in (b) displays fluctuations of the number of transactions around January 6, 2023, highlighting a significant surge in activity for the aforementioned collections during this period.}
\label{Cap_N}
\end{figure}

Capitalization $K(t)$ and the number of transactions $N_{\Delta t=24h}(t)$ for all the collections are depicted in Fig.~\ref{Cap_N}(a)(b). As shown in panel (a), most collections experienced a decrease in $K(t)$ during the studied period, reflecting the overall downtrend in the NFT market. An exception to this trend is Milady Maker collection, which experienced significant capitalization gains in April and May 2023. These gains were accompanied by a surge in transaction activity, with the collection recording the highest daily transaction count among all collections in the period studied, reaching approximately 1,600 transactions on May 10, 2023. The unique trend observed in Milady Maker NFTs can be attributed to Elon Musk's tweet about this collection on this specific day and to the subsequent price rally~\cite{miladyMusk}. Other collections with a lower capitalization that experienced slight valuation increases include Lil Pudgys, Sappy Seals, and Pudgy Penguins. These collections also rank among the most frequently traded ones, although the general 
rule is that the most capitalized collections tend to be traded more frequently. The only exception is CryptoPunks, which despite the largest capitalization, have seen fewer transactions recently, likely due to their long-standing presence in the market. Another interesting spike in activity was observed on January 6, 2023 among less frequently traded and lower-capitalized collections such as Alien Frens, Killer GFs, Rug Radio - Genesis, Karafuru, Coolman's Universe, 3Landers, and Goblintown. On that date, these collections experienced a significantly higher than average transaction volume and a notable drop in capitalization, as illustrated in inset to Fig.~\ref{Cap_N}(b).


\begin{figure}
\includegraphics[width=0.49\textwidth]{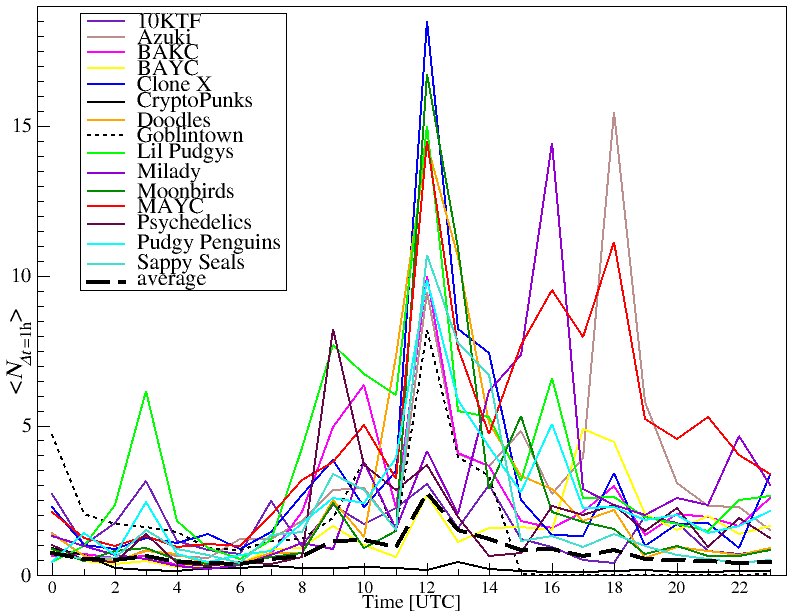}
\caption{Daily pattern of the number of transactions in 1 hour interval averaged over all trading days $\langle N_{\Delta t=1h} \rangle$.}
\label{24Nav}
\end{figure}

Various patterns of activity in financial markets are observed~\cite{AndersenTG-1998a,AndersenTG-1999a,BollerslevT-2000a}, which can be linked to the start/end of trading sessions, specific days of the week, and macroeconomic news announcements. This patterned behavior is mirrored in the cryptocurrency market as well~\cite{Watorek2023chaos}. Given this context, it became compelling to investigate the presence of similar trading activity patterns within the NFT market. By segmenting the number of transactions time series $N_{\Delta t =1h}$ into 500 time series for 24-hour-long periods and averaging across all trading days, a daily transaction pattern was identified - see Fig.~\ref{24Nav}. A period of the most frequent trading corresponds to the US daytime trading hours peaking around 12:00 UTC, which is 8:00 a.m. in New York. Most of the liquid collections see their highest activity levels during this window. Additionally, two other notable peaks occur at 16:00 and 18:00 UTC, when the most frequently traded collection - Monkey Ape Yacht Club - had the most transactions on average.


\begin{figure}
\includegraphics[width=0.49\textwidth]{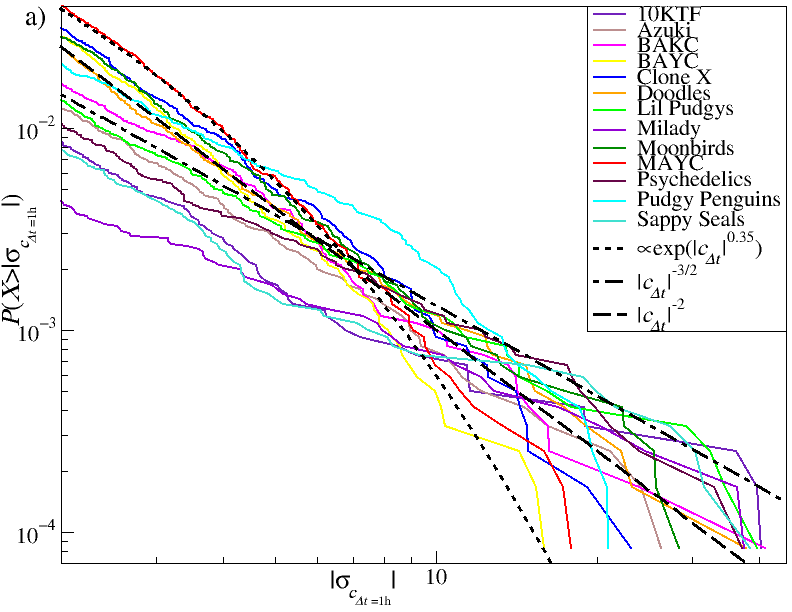}
\includegraphics[width=0.49\textwidth]{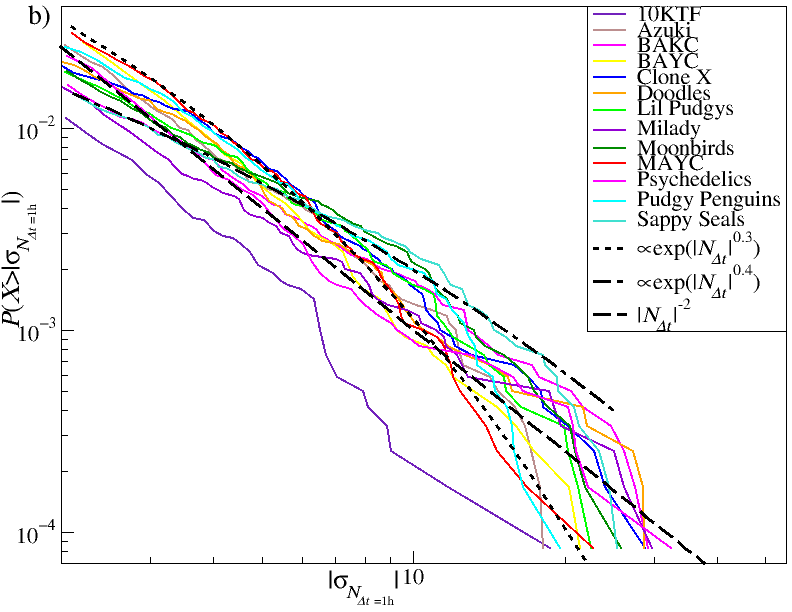}
\caption{Complementary cumulative distribution functions for (a) absolute logarithmic increments of collection capitalization expressed in USD $|c_{\Delta t=1h}(t)|$ and (b) the number of transactions aggregated hourly $N_{\Delta t=1h}(t)$.}
\label{Rozklady1h}
\end{figure}

Understanding the characteristics of time series is crucial before proceeding with a correlation analysis, because this knowledge can help one in choosing the right correlation metrics. For this purpose, cumulative distributions of the capitalization increments $c_{\Delta t}$ and the number of transactions $N_{\Delta t}$ were analyzed for the 13 most liquid collections (those with $\%0_{\Delta t=1{\rm h}} < 50\%$ in Table~\ref{NFTnames}). Fig.~\ref{Rozklady1h} shows that the complementary cumulative distribution functions (CCDFs) $P(X>\sigma)$, where $\sigma$ denotes the estimate of standard deviation, of $|c_{\Delta t}|$ and $N_{\Delta t}$ for all the collections exhibit heavy tails -- a common characteristics of financial time series~\cite{Ausloos2000,ContR-2001a,Plerou2004,Watorek2021distr}.

For the absolute capitalization increments $|c_{\Delta t}|$, the CCDF tails follow a power law $x^{-\gamma}$ with the exponent $\gamma$ varying between 1.5 and 2.0 for most collections -- see Fig.~\ref{Rozklady1h}(a). For the most liquid ones, such as Bored Ape Yacht Club and Mutant Ape Yacht Club, a stretched exponential function ${\rm exp}(x^\beta)$ provides a good fit if $\beta \approx 0.35$. In terms of the number of transactions $N_{\Delta t}$, the CCDF tails are less heavy as compared to those for $|c_{\Delta t}|$, with the stretched exponential function accurately fitting the data for $0.3 \le \beta \le 0.4$. These findings align closely with the results from a previous study on the most liquid collections from the Solana blockchain~\cite{Szydlo2024}, where similar patterns were observed in the distributions of absolute capitalization increments and the number of transactions.


\begin{figure}
\includegraphics[width=0.49\textwidth]{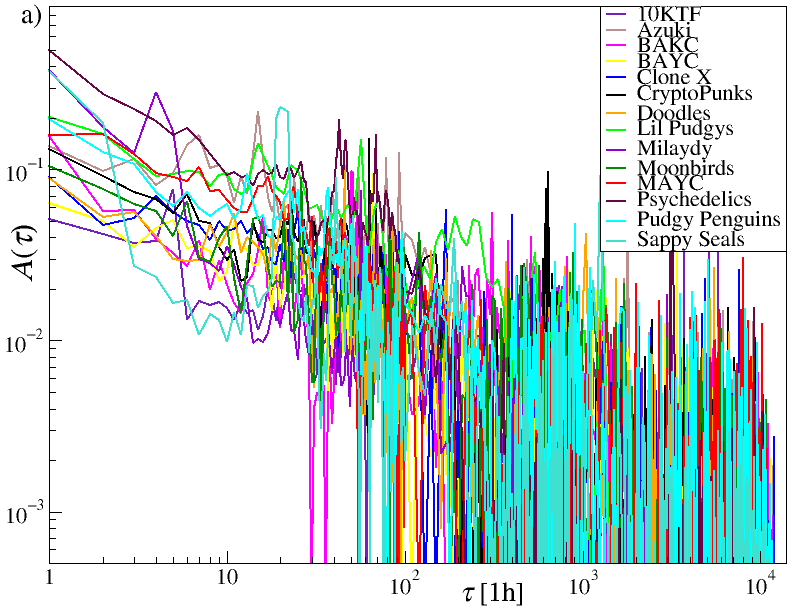}
\includegraphics[width=0.49\textwidth]{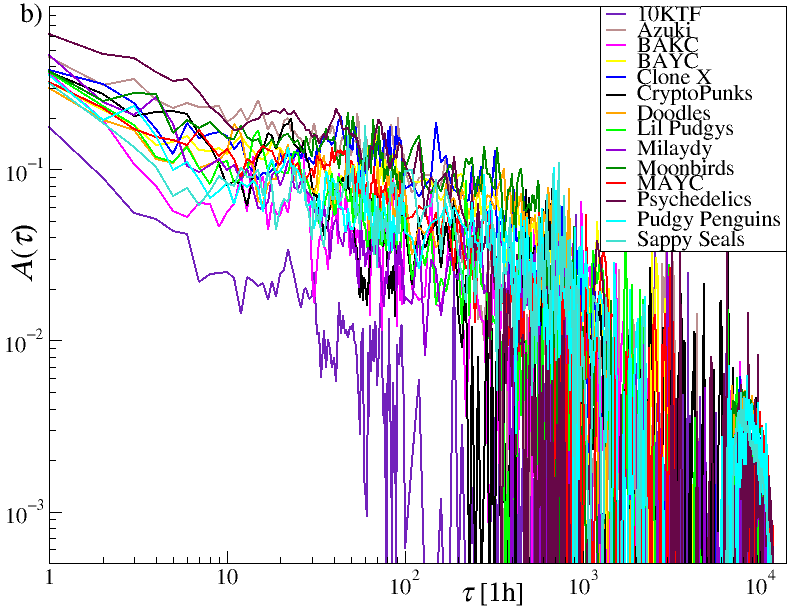}
\caption{Autocorrelation functions of $c_{\Delta t=1h}$ (a) and $N_{\Delta t=1h}$ (b) for all the collections.}
\label{ACF_KN}
\end{figure}

Another measure of temporal relationships in time series is the autocorrelation function
\begin{equation}
A(x,\Delta i) = { 1/T \sum_{i=1}^T \left[ x(i) - \langle x(i) \rangle_i \right] \left[ x(i+\Delta i) - \langle x(i) \rangle_i \right] \over \sigma^2_x},
\end{equation}
where $\sigma_x^2$ denotes the estimate of variance of a time series ${x(i)}$, $\langle \cdot \rangle$ is the estimate of mean, and $\Delta i$ refers to a temporal lag measured in data points (which can be translated to time by $\tau = \Delta i \Delta t$). For the financial logarithmic returns, the autocorrelation function (ACF) shows a characteristic behavior of dropping immediately to zero if their sign is taken into consideration and slowly decreasing according to a power law if their absolute values are considered~\cite{Gopikrishnan1999}. Fig.~\ref{ACF_KN} presents ACFs for the 13 most liquid collections for both $|c_{\Delta t}|$ and $N_{\Delta t}$ with $\Delta t=1$h. A consistent observation across these collections is a slow decay of \(A(x,\tau)\) signaling the presence of long-range memory. For certain collections, a power-law decay can also be observed. These long-range autocorrelations persist for several hundred hours. A similar power-law decay in ACF was observed in a prior study~\cite{Szydlo2024} for the most liquid collections on the Solana blockchain.


\subsection{Detrended correlations}

Given that the properties of the time series discussed in section~\ref{Dataspec} were identified as heavy-tailed with a long memory, the detrended cross-correlation coefficient $\rho(q,s)$~\cite{kwapien2015} was employed alongside the standard Pearson coefficient to compute correlations among the characteristics of the considered collections. The method used to derive $\rho(q,s)$ is the multifractal version of detrended fluctuation analysis (MFDFA)~\cite{PengCK-1994a,KantelhardtJ-2002a}, which was designed to identify long-range power-law auto- and cross-correlations that give rise to trends across various time scales. Consequently, in contrast to the conventional trend-elimination techniques, MFDFA and the related measures such as the $q$-dependent detrended cross-correlation coefficient $\rho(q,s)$ are effective in managing nonstationarity~\cite{JiangZQ-2019a}. The coefficient $\rho(q,s)$ allows one for an examination of cross-correlation intensity across various time scales. The parameter $q$ enables for the correlation analysis to focus on a specific range of fluctuation magnitudes.

The procedure for calculating the $\rho(q,s)$ coefficient consists of the following steps. Two time series $\{x(i)\}_{i=1,...,T}$ and $\{y(i)\}_{i=1,...,T}$ of length $T$ are divided into $M_s$ segments of length $s$ starting from its opposite ends. Then the time series are integrated, and in each segment $\nu$, the polynomial trend is removed:
\begin{align}
X_{\nu}(s,i) = \sum_{j=1}^i x(\nu s + j) - P^{(m)}_{X,s,\nu}(i), \\
Y_{\nu}(s,i) = \sum_{j=1}^i y(\nu s + j) - P^{(m)}_{Y,s,\nu}(i),
\end{align}
by using ordinary least-squares method and the polynomials $P^{(m)}$ of order $m$ are utilized. Specifically, an order of $m=2$ was chosen for its effectiveness with financial time series~\cite{OswiecimkaP-2013a}. Following this procedure, a total of $2 M_s$ segments containing detrended signals are generated. Subsequently, the estimate of variance and estimate of covariance for each segment $\nu$ are calculated as follows:
\begin{align}
f^2_{\rm ZZ} (s,\nu) = \frac{1}{s}\sum_{i=1}^s (Z_{\nu}(s,i) )^2, \\
f^2_{\rm XY} (s,\nu) = \frac{1}{s}\sum_{i=1}^s X_{\nu}(s,i)\times Y_{\nu}(s,i),
\end{align}
where $Z$ means $X$ or $Y$. These quantities are used to calculate a family of the fluctuation functions:
\begin{align}
F_{\rm ZZ} (q,s) = {1 \over 2 M_s} \sum_{\nu=0}^{2 M_s-1} \left[ f^2_{\rm ZZ} (s,\nu)\right]^{q/2},
\label{eq::fq.zz} \\
F_{\rm XY} (q,s) = {1 \over 2 M_s} \sum_{\nu=0}^{2 M_s-1} \textrm{sign} \left[ f^2_{\rm XY}(s,\nu)\right] |f^2_{\rm XY} (s,\nu)|^{q/2},
\label{eq::fq.xy}
\end{align}
where a sign function $\textrm{sign} \left[ f^2_{\rm XY}(s,\nu)\right]$ is maintained to ensure that outcomes remain consistent across various values of $q$s.

Finally, the formula for the  $q$-dependent detrended correlation coefficient is presented as follows~\cite{kwapien2015}:

\begin{equation}
\rho^{\rm XY}(q,s) = {F_{\rm XY}(q,s) \over \sqrt{F_{\rm XX}(q,s) F_{\rm YY}(q,s)}}.
\label{eq::rhoq}
\end{equation}

For $q=2$, the given definition can be interpreted as a detrended version of the Pearson cross-correlation coefficient $C$~\cite{ZebendeG-2011a}. In this context, the parameter $q$ functions as a selective mechanism that either suppress ($q<2$) or amplifies ($q>2$) the fluctuation variance/covariance determined within the segments $\nu$ in Eqs.~(\ref{eq::fq.zz}) and (\ref{eq::fq.xy})) When $q<2$, segments characterized by small fluctuations have a greater influence on $\rho_q(s)$, whereas, for $q>2$, segments marked by large fluctuations have a heightened impact. Hence, utilizing this measure allows for the identification of the range of fluctuation sizes contributing to the noted correlations.

The methods presented above can also be used to investigate the presence of multifractality for a single time series. If the bivariate or univariate fluctuation functions can be approximated by a power-law relation
\begin{equation}
F(q,s) \sim s^{h(q)},
\label{eq::fractal.scaling}
\end{equation}
where $h(q)$ represents a non-increasing function of $q$, known as the generalized Hurst exponent~\cite{BarabasiAL-1991a}. This function indicates that the analyzed time series exhibits a fractal structure. When $h(q)$ remains constant, the structure is identified as monofractal, with $h(q)=H$ being the Hurst exponent - a metric for time series persistence. If $h(q)$ is not constant, the structure is recognized as multifractal. The source of multifractality in the time series is the occurrence of non-linear correlations~\cite{DrozdzS-2009a,zhou2009components,RAK2018,KwapienJ-2023a}, which are related to the existence of long memory manifested by power-law decay of the autocorrelation function. As observed in Fig.~\ref{ACF_KN}, such a power-law decay of the autocorrelation function occurs in the case of the considered collection time series. The multifractality of the financial time series is also a stylized fact~\cite{JiangZQ-2019a,Kutner2019}. In our previous work~\cite{Szydlo2024}, the sign of multifractality of the NFT collections from the Solana blockchain time series characteristics was observed for some cases, especially for large fluctuations. 

\begin{figure}
\includegraphics[width=0.49\textwidth]{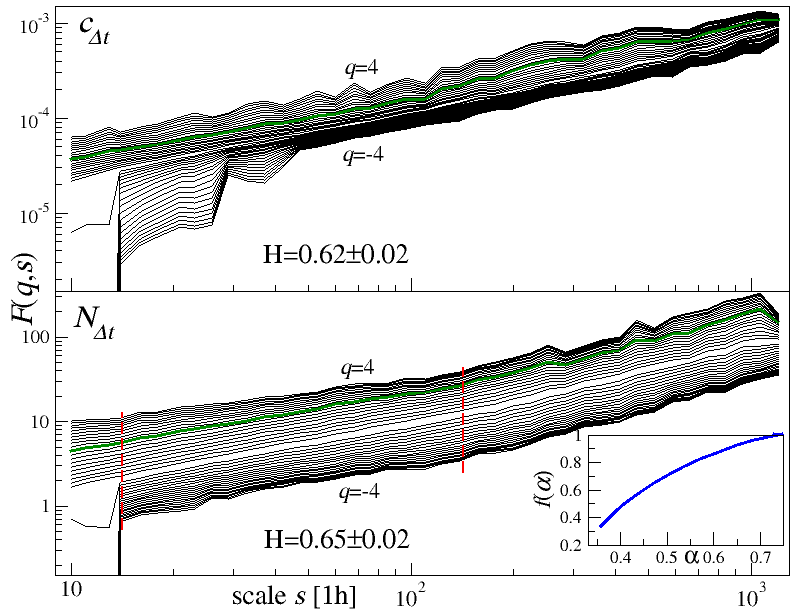}
\caption{Fluctuation functions $F(q,s)$ with $q \in [-4,4]$ calculated for the number of transactions $N_{\Delta t=1h}$ and logarithmic increments of the total collection capitalization $c_{\Delta t=1h}$ for Mutant Ape Yacht Club collection. In main panels, thick green lines represent $F(q=2,s)$, which are utilized in the estimation of the Hurst exponent $H$ with the standard error. Vertical red dashed lines point out to a scale range where the family of $F(q,s)$ exhibit power-law dependence for different values of $q$, from which the singularity spectrum $f(\alpha)$ can be calculated (insets).}
\label{FqMAYC}
\end{figure}

This study investigates the fractal characteristics of capitalization increments $c_{\Delta t=1h}$ and number of transactions $N_{\Delta t=1h}$ for the most liquid collection during the examined period -- Mutant Ape Yacht Club. Fig.~\ref{FqMAYC} presents $F(q,s)$ for both $c_{\Delta t}$  (top) and $N_{\Delta t}$ (bottom) time series in double-logarithmicscale with $-4 \le q \le 4$ and $10 \le s \le 1,200$. As can be seen in both cases, there is at least a decade-long scaling range long enough to determine the slope of $F(q=2,s)$ (green line in Fig.~\ref{FqMAYC}) and, thus, to obtain the Hurst exponent $H \equiv h(q=2)$ according to Eq.~(\ref{eq::fractal.scaling}). The values indicate that both the analyzed time series are persistent with $H=0.62\pm0.02$ for $c_{\Delta t}$ and $H=0.65\pm0.02$ for $N_{\Delta t}$. These findings are in agreement with the results from our previous study based on NFT collections on the Solana blockchain~\cite{Szydlo2024}. However, when examining $h(q)$ across a range of $q \in [-4,4]$, a sufficient scaling region is identifiable solely for the number of transactions (as it is highlighted by red dashed lines in the bottom panel of Fig.~\ref{FqMAYC}). For the capitalization increments, the scaling region is noticeable only for negative $q$s, and it does not extend beyond $q>2$. Having calculated $h(q)$, it is straightforward to derive the singularity spectra $f(\alpha)$ in accordance to the following formulas:
\begin{equation}
\alpha = h(q) + q {dh(q) \over dq}, \quad f(\alpha) = q(\alpha - h(q)) + 1.
\label{eq::singularity spectra}
\end{equation}
The multifractal spectrum for the number of transactions $N_{\Delta t}$ is depicted in the inset of Fig.~\ref{FqMAYC}. Notably, the spectrum exhibits only the left branch indicating that the multifractal structure is evident only for medium and large fluctuations. For both $c_{\Delta t}$ and $N_{\Delta t}$, the negative $q$s that are associated with smaller fluctuations exhibit a monofractal regime indicated by the absence of a right arm in $f(\alpha)$. This left-side asymmetry in the singularity spectrum is commonly observed in financial time series where smaller fluctuations tend to represent noise, while the medium and large ones contain genuine information~\cite{drozdz2015mf,DrozdzS-2018a,JiangZQ-2019a,KwapienJ-2022a,watorekfutnet2022,WatorekM-2023b,Brouty2024}.


\subsection{Correlation matrix}

In order to study correlations between the time series representing $I=90$ collections: the logarithmic increments of total collection capitalization $c_{\Delta t}$ and the number of transactions $N_{\Delta t}$, the correlation matrices ${\bf C}$ and ${\bf C}^{\rho}(q,s)$ based on the Pearson correlation coefficient~\cite{Pearson1895} $P_{ij}$ and the $q$-dependent detrended correlation coefficient~\cite{kwapien2015} $\rho_{ij}(q,s)$ were created, respectively. After diagonalization of ${\bf C}$ and ${\bf C}^{\rho}$:
\begin{eqnarray}
{\bf C}{\bf v}_i = \lambda_i {\bf v}_i,\\
{\bf C}^{\rho}(q,s){\bf v}_i^{\rho}(q,s) = \lambda_i^{\rho}(q,s) {\bf v}_i^{\rho}(q,s),
\end{eqnarray}
its eigenvalues $\lambda_i$, $\lambda^{\rho}$ and eigenvectors ${\bf v}_i=\{v_{ij}\}$, ${\bf v}_i^{\rho}=\{v_{ij}^{\rho}\}$ are derived for $i,j=1,...,I$. The resulting empirical eigenvalue distribution can be compared with the Marchenko-Pastur distribution linked to the Wishart ensemble~\cite{randommatrix} of random matrices ${\bf W}$. This ensemble exemplifies the inherent characteristics of uncorrelated independent and identically distributed (i.i.d.) random variables following a Gaussian distribution, $N(0,\sigma)$~\cite{wishart1928}. The probability density function that describes the eigenvalue distribution of ${\bf W}$ is given by~\cite{Marchenko1967}
\begin{eqnarray}
\nonumber
\phi_\textrm{W}(\lambda)={1 \over I} \sum_{i=1}^N \delta(\lambda - \lambda_k) = {Q \over 2 \pi \sigma_{\textrm{W}}^2} {\sqrt{(\lambda_{+}-\lambda)(\lambda-\lambda_{-})} \over \lambda},\\
\label{rhoW}
\lambda_{\pm} = \sigma_{\textrm{W}}^2 (1 + 1/Q \pm 2 \sqrt{1 \over Q}),
\end{eqnarray}
where $\lambda\in[\lambda_-,\lambda_+]$ and $Q=T/I$. Here, $T$ and $I$ denote the number of time series and their length, respectively. This relationship is strictly valid in the limit $T, K \to \infty$. However, contrasting an empirical eigenvalue distribution with the Marchenko-Pastur distribution helps in identifying the presence of any correlated patterns within the data.

\section{Correlations between collection characteristics}

\subsection{Eigenvalues and off-diagonal elements distributions}
\label{cceig}

Properties of the correlation matrices derived from the time series of total collection capitalization increments $c_{\Delta t}$ and the number of transactions $N_{\Delta t}$ were analyzed and compared with theoretical predictions from the random matrix theory~\cite{randommatrix}. The empirical eigenvalue distributions were compared with the Marchenko–Pastur distribution corresponding to the Wishart random matrix ensemble~\cite{wishart1928,Marchenko1967} that represents the universal properties of uncorrelated i.i.d. random variables with a Gaussian distribution $N(0,1)$. Moreover, the study included an examination of how the off-diagonal elements in the correlation matrices relate to a Gaussian distribution, enhancing the scope of the analysis.

\begin{figure}
\centering
\includegraphics[width=0.495\textwidth]{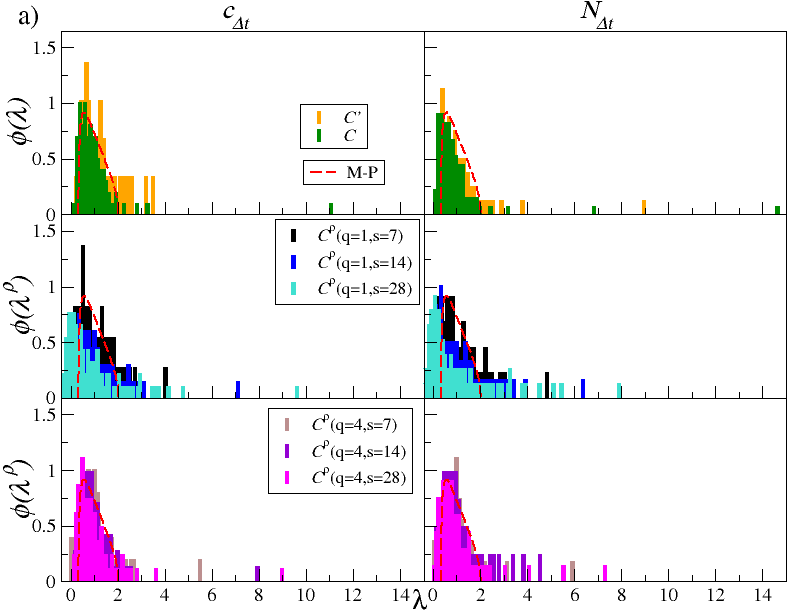}
\includegraphics[width=0.495\textwidth]{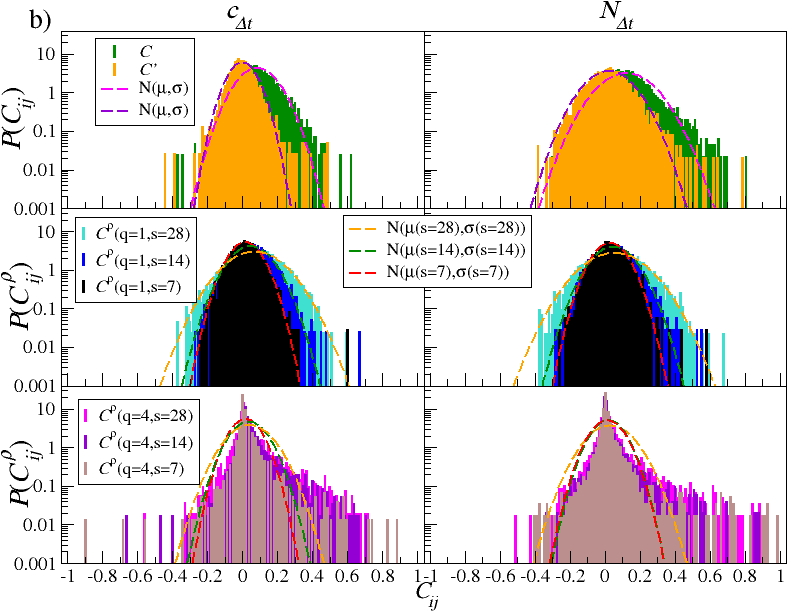}
\caption{(a) Eigenvalue distributions of the correlation matrices ${\bf C}$, ${\bf C}^{\rho}(q=1,s)$ and ${\bf C}^{\rho}(q=4,s)$ (top, middle, bottom) created from the total collection capitalization increments $c_{\Delta t}$ (left) and the number of transactions $N_{\Delta t}$ (right) for $\Delta t=1$h. The theoretical Marchenko–Pastur distribution $\phi_{\rm W}(\lambda)$ is marked by dashed red lines. (b) Probability distribution functions (histograms) of the off-diagonal elements of the same correlation matrices and data as in (a). Dashed lines show a fitted normal distribution to each empirical distribution.}
\label{fig::cc}
\end{figure}

In the Pearson-coefficient approach (upper panels of Fig.~\ref{fig::cc}(a)), the number of eigenvalues exceeds the bounds of the Marchenko-Pastur (M-P) region, which contains the bulk of eigenvalues corresponding to random fluctuations. This fact points out to the presence of statistically significant correlations among the analyzed time series. Particularly noteworthy is the largest eigenvalue $\lambda_1$, which stands well outside this region. This is a common phenomenon in the financial markets, where a typical eigenvalue spectrum of the Pearson-coefficient based correlation matrices consists of a large $\lambda_1$, separated from the remaining ones by a considerable gap, that represents a collective market factor representing all the assets evolving in concert~\cite{Plerou2002,KwapienJ-2012a}.

Mature financial markets are also characterized by the existence of a few other elevated non-random eigenvalues that correspond to market sectors (e.g., industries in the case of stock markets or geographical regions in the case of currencies)~\cite{McDonaldM-2005a,KwapienJ-2012a}. Regarding both $c_{\Delta t}$ and $N_{\Delta t}$, it is observed that the eigenvalues $\lambda_2$ through $\lambda_4$ deviate from the M-P universal prediction, which is particularly noteworthy for $\lambda_2$ in the case of $N_{\Delta t}$. This suggests that the time series of the number of transactions are more closely correlated within groups than those for the capitalization increments. The significance of $\lambda_2$ becomes even more evident once the variance contribution associated with $\lambda_1$ has been filtered out by using a standard procedure~\cite{Plerou2002,KWAPIEN2006} (see top panels of Fig.~\ref{fig::cc}(a)).

The market factor becomes weaker, although still significant, if the matrices ${\bf C}^{\rho}(q=1,s)$ and ${\bf C}^{\rho}(q=4,s)$ are considered (middle and bottom panels of Fig~\ref{fig::cc}(a)). In this case, one could also observe an effect of the correlations increasing with time scale $s$, which is typical for the financial markets~\cite{Watorek2019a}. The probability distribution function (histograms) of the off-diagonal elements of the correlation matrices ${\bf C}$ and ${\bf C}^{\rho}(q,s)$ offers complementary information to the eigenvalue distribution. Deviations from a Gaussian distribution in this case are visible in all the cases in Fig~\ref{fig::cc}(b). The fattest tails are observed for $q=4$ (large fluctuations) for $N_{\Delta t}$ - see bottom left panel of Fig~\ref{fig::cc}(b), where the extreme elements $C_{ij}^{\rho}(q=4,s)$ exceed 0.9 even on the shortest time scale $s=7$ days. These large correlations can be traced back to a number of transactions surge across various less-traded collections on the same day Jan, 6 2023 (inset of Fig.~\ref{Cap_N}(b)). Their origin will be explained in the following section.


\subsection{Eigenvector components}
\label{eigcomp}

\begin{figure}
\includegraphics[width=0.49\textwidth]{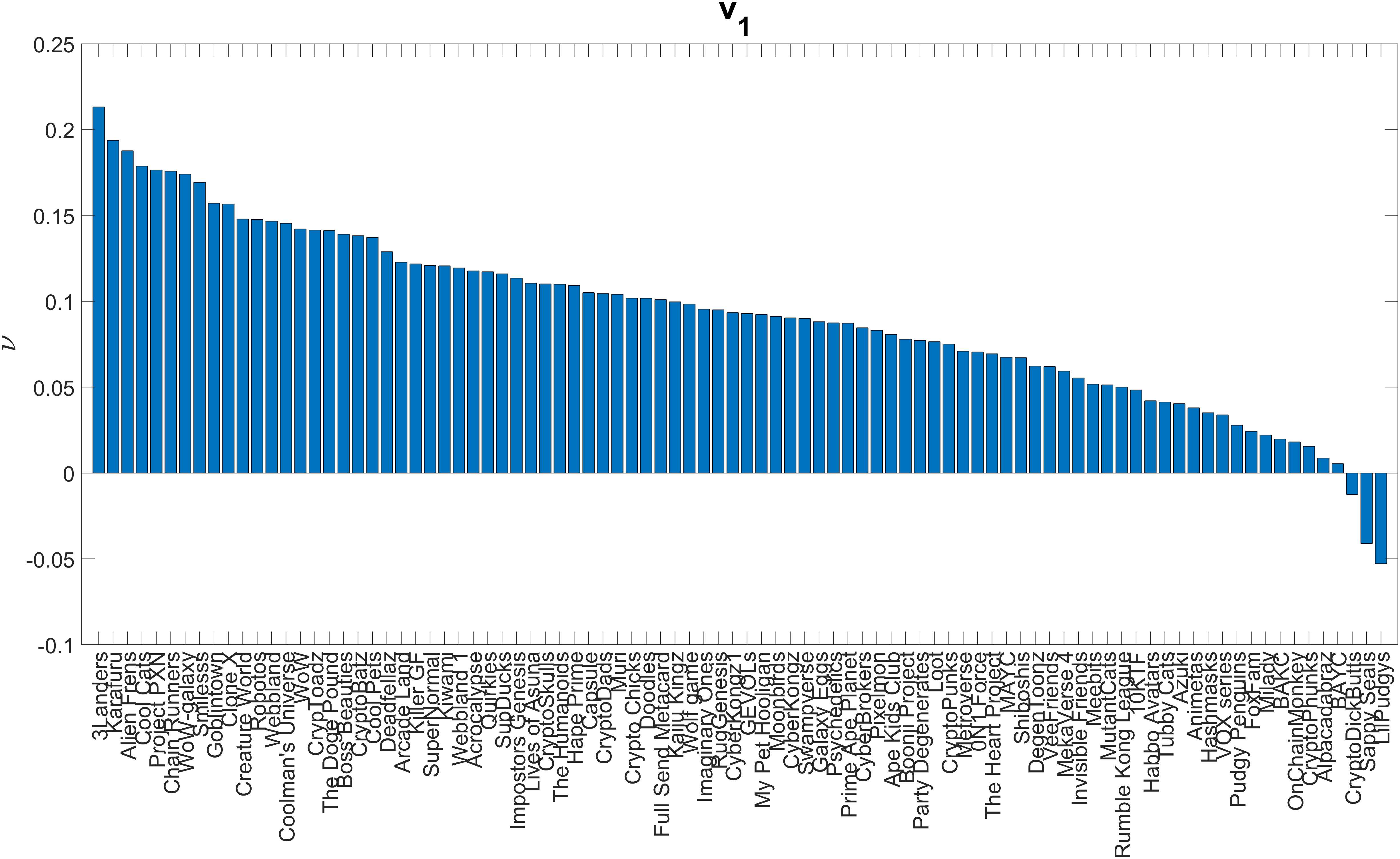}
\includegraphics[width=0.49\textwidth]{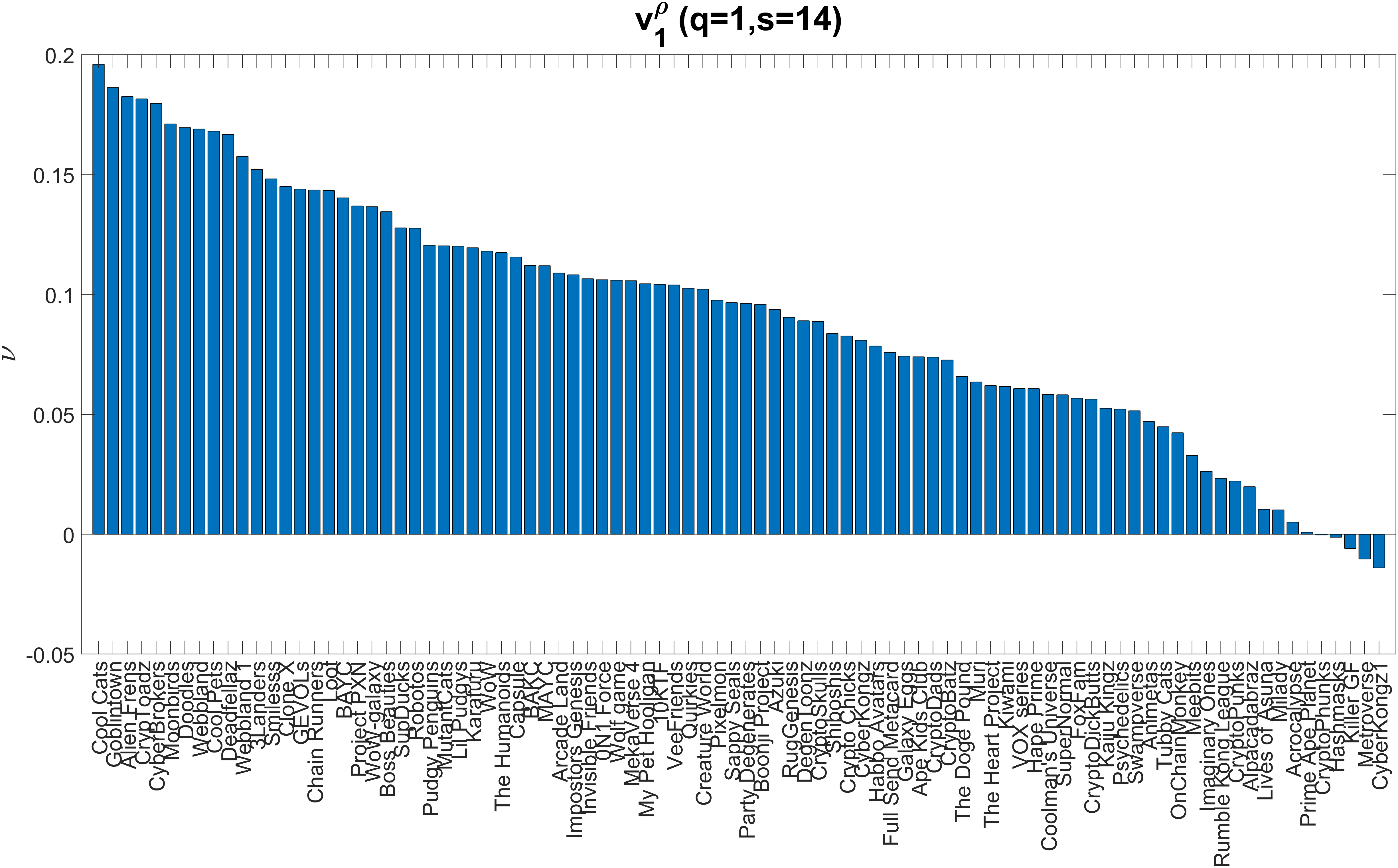}
\includegraphics[width=0.49\textwidth]{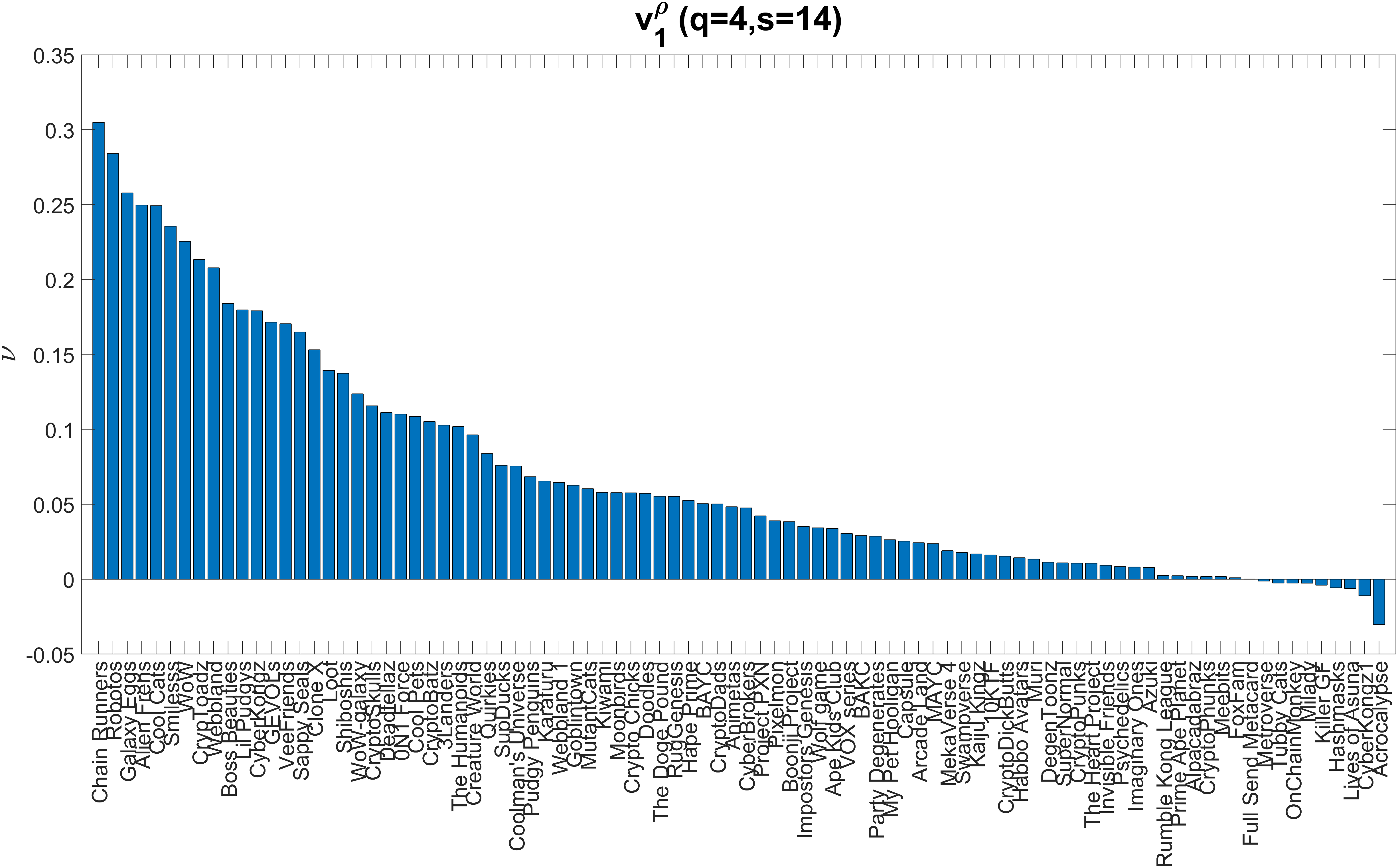}
\caption{Expansion coefficients of the eigenvector ${\bf v}_1$ associated with largest eigenvalue $\lambda_1$ of the Pearson correlation matrix ${\bf C}$ (top) and of the eigenvectors ${\bf v}_1^{\rho}(q=1,s=14)$ and ${\bf v}_1^{\rho}(q=4,s=14)$ associated with largest eigenvalue $\lambda_1^{\rho}(q=1,s=14)$ (middle) and $\lambda_1^{\rho}(q=4,s=14)$ (bottom), respectively. All correlation matrices were calculated based on the total capitalization increment time series $c_{\Delta t}$.}
\label{V1_Cap}
\end{figure}

By examining the expansion coefficients $v_{ij}$ of the eigenvectors ${\bf v}_i$ associated with the eigenvalues $\lambda_i$ that stand out significantly from the M-P region, it is possible to estimate their contributions by using either the Pearson or the detrended correlation coefficients derived from the time series of total collection capitalization increments $c_{\Delta t}$ and the number of transactions $N_{\Delta t}$ for $\lambda_1$ and $\lambda_2$. In the case of $c_{\Delta t}$ and the Pearson correlation coefficient (top panel of Fig.~\ref{V1_Cap}), the collective character of ${\bf v}_1$ is distorted by the existence of the significant expansion coefficients $v_{1j}$ with an opposite sign associated to: Lil Pudgys and Sappy Seals ($v_{1j}\approx -0.05$). On the other hand, the highest contribution to ${\bf v}_1$ comes from 3Landers, Karafuru and Alien Frens ($v_{1j} \in [0.18, 0.22]$). On the one hand, for $q=1$ that corresponds to fluctuations of all magnitudes (Fig.~\ref{V1_Cap}, middle panel), the negative coefficients correspond to CyberKongz1, Metroverse and Killer GF collections, while for $q=4$ that amplifies large fluctuations (Fig.~\ref{V1_Cap}, bottom panel), the negative expansion coefficients correspond to Acrocalypse and CyberKongz1 collections. On the other hand, the greatest contribution to ${\bf v}_1^{\rho}(q,s)$ is made by the expansion coefficients associated with Cool Cats, Goblintown, Alien Frens, and CrypToadz collections ($v_{1j}^{\rho} \in [0.18,0.2]$) for $q=1$ and with Chain Runners, Robotos, Galaxy Eggs, Alien Frens and Cool Cats collections ($v_{1j}^{\rho} \in [0.25,0.31]$) for $q=4$. It is worth noting that the largest $v_{1j}^{\rho}$ are smaller for $q=1$ than for $q=4$ and for ${\bf C}$.


A general conclusion that can be derived from the eigenvectors ${\bf v}_1$ and ${\bf v}_1^{\rho}$ is that the most substantial contributions to the market factor do not originate from the most liquid and highly capitalized collections, such as Mutant Ape Yacht Club, Bored Ape Yacht Club, or CryptoPunks, but from less liquid collections with lower capitalization. This phenomenon can be attributed to the generally weak average correlation and, thus, to relatively minor significance of the market factor, as evidenced by the low values of the largest eigenvalue $\lambda_1$ and $\lambda_1^{\rho}$ in relation to other financial markets~\cite{KwapienJ-2012a}, including the cryptocurrency market~\cite{entropy2021b}. In fact, the source of the significant matrix elements was singular, exceptionally large synchronous events that occurred among less-liquid collections, like the activity rush on January 6, 2023.

\begin{figure}
\includegraphics[width=0.49\textwidth]{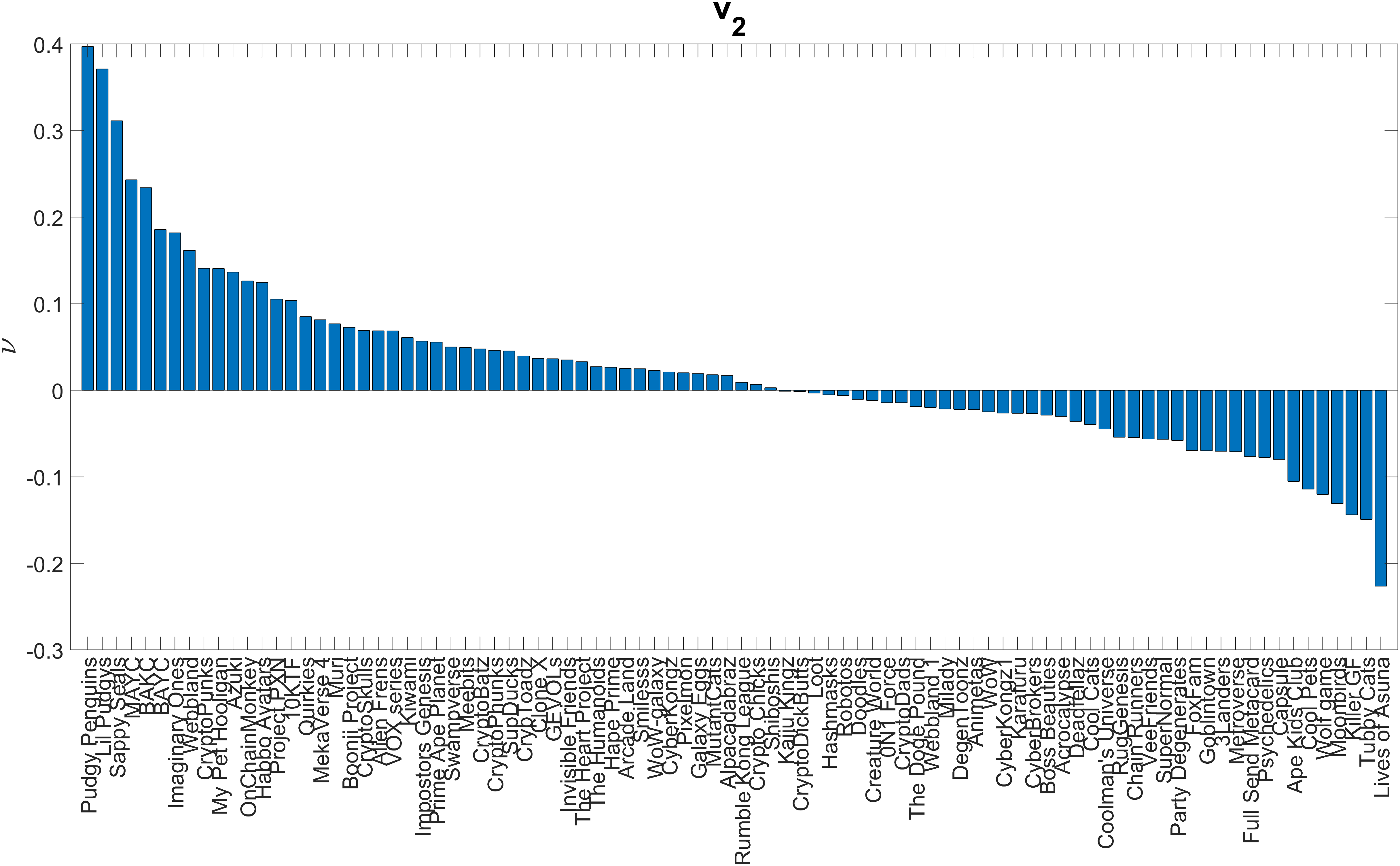}
\includegraphics[width=0.49\textwidth]{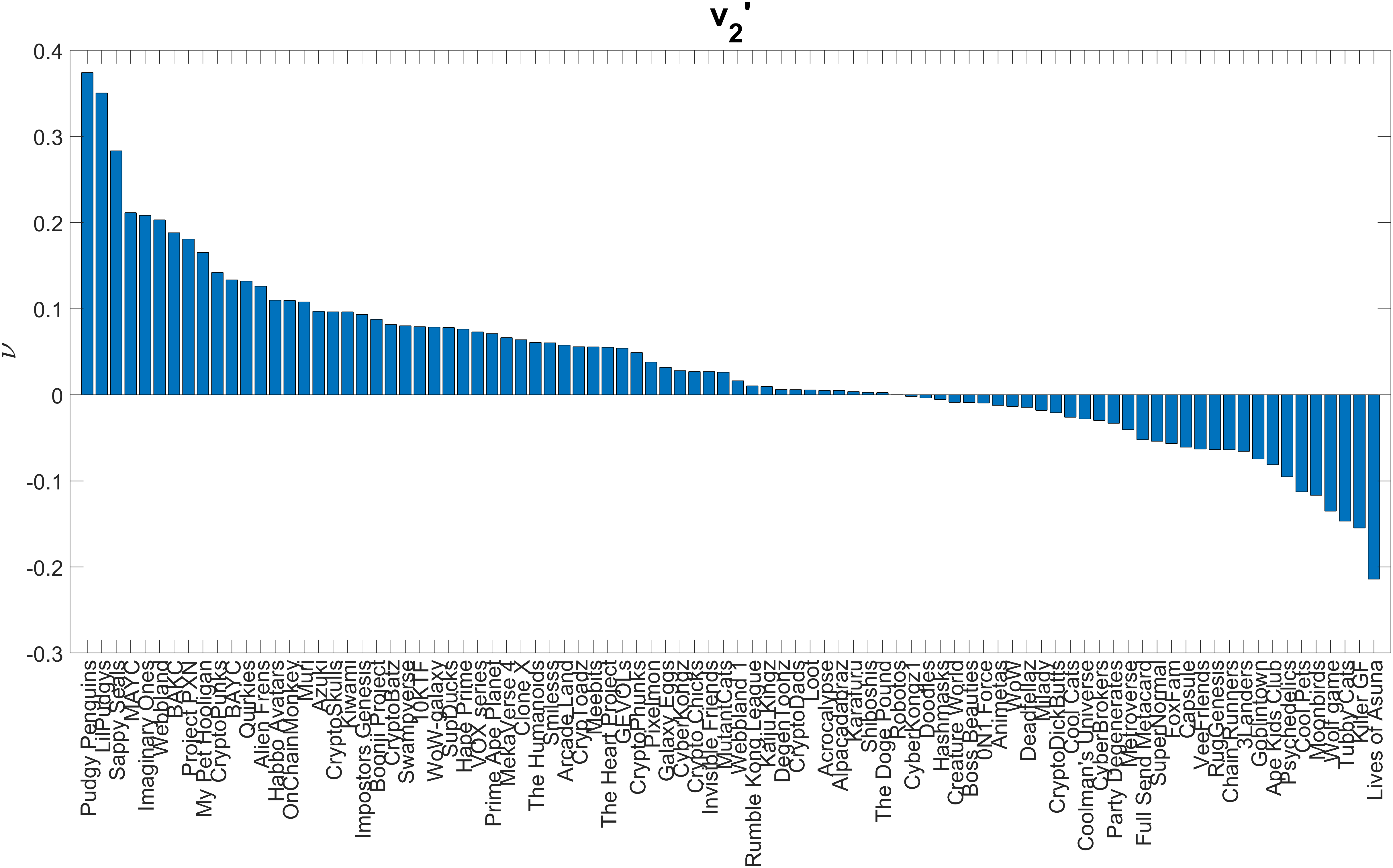}
\includegraphics[width=0.49\textwidth]{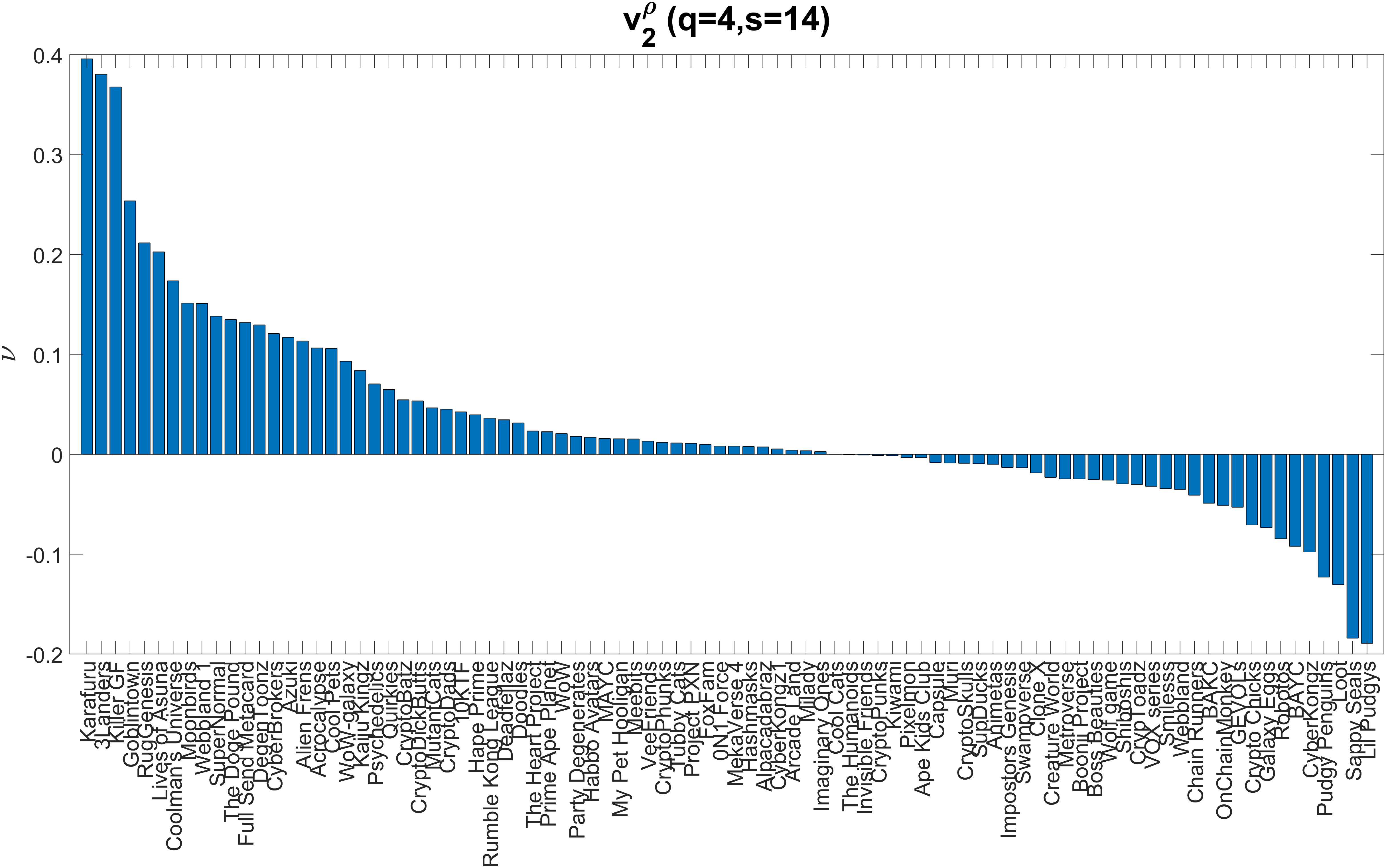}
\caption{Expansion coefficients of the eigenvector ${\bf v}_2$ associated with the 2nd largest eigenvalue $\lambda_2$ of the Pearson correlation matrix ${\bf C}$ (top), the eigenvector ${\bf v}_2'$ associated with the largest eigenvalue $\lambda_2'$ of the filtered correlation matrix ${\bf C}'$ (middle), and the eigenvector ${\bf v}_2^{\rho}$ associated with the largest eigenvalue $\lambda_2^{\rho}$ of the detrended correlation matrix ${\bf C}^{\rho}(q=4,s=14)$ (bottom). All correlation matrices were calculated based on the total capitalization increment time series $c_{\Delta t}$.}
\label{V2_Cap}
\end{figure}

In order to identify collections that contribute substantially to the eigenvector ${\bf v}_2$, it is recommended to remove that part of the signal variance that corresponds to the largest eigenvalue $\lambda_1$. It can be done by using a regression-based method~\cite{Plerou2002,KwapienJ-2012a}:
\begin{eqnarray}
\nonumber
c_{\Delta t}^{^{\rm (i)}}(k) = a^{^{\rm (i)}} + b^{^{\rm (i)}} Z_1(k) + \epsilon^{^{\rm (i)}}(k),\\
Z_1(k) = \sum_{m=1}^I v_{1m} c_{\Delta t}^{(m)}(k),
\label{eq::regression}
\end{eqnarray}
where $Z_1(k)$ is the contribution to total variance associated with $\lambda_1$ ($k=1,...,T$), and diagonalizing the filtered matrix ${\bf C}'$ constructed from the residual time series $\epsilon^{^{(i)}}(k)$ ($i=1,...,I$). The corresponding eigenvalue-eigenvector pairs for such a filtered matrix have an index decreased by unity $\lambda_i \rightarrow \lambda_{i-1}'$, but for the sake of simplicity, we shall leave the index unchanged: $\lambda_2', {\bf v}_2'$. Fig.~\ref{V2_Cap} shows the expansion coefficients $v_{2j}$ for the original matrix ${\bf C}$ without applying Eqs.~(\ref{eq::regression}) (top), for the filtered matrix ${\bf C}'$ (middle), and for the detrended correlation matrix ${\bf C}^{\rho}(q=4,s=14)$ after filtering analogous to Eqs.~(\ref{eq::regression}) (bottom).

The first conclusion that can be drawn from the coefficients $v_{2j}$ is that Pudgy Penguins and the collection originated from it - Lil Pudgys, that were created by the same producer - TheIglooCompany, show strong correlations in terms of $c_{\Delta t}$. The same applies to ape-themed collections (Bored Ape Yacht Club, Bored Ape Kennel Club, and Mutant Ape Yacht Club) released by Yuga Labs. The related expansion coefficients are among the largest ones ($v_{2j} \in [0.2,0.4]$), together with Sappy Seals (top panel of Fig.~\ref{V2_Cap}). The close relations among collections released by TheIglooCompany become even more pronounced after removing the contribution of $\lambda_1$ (middle panel of Fig.~\ref{V2_Cap}), where their expansion coefficients $v_{2j}' \approx0.38$ are much larger than for the other collections. The largest opposite-sign contribution to ${\bf v}_2'$, which indicates a different behavior, is observed for Lives of Asuna collection (middle panel of Fig.~\ref{V2_Cap}). Other relations between collections can be inferred from ${\bf v}_2^{\rho}(q=4,s=14)$ (bottom panel of Fig.~\ref{V2_Cap}) for $q=4$, which amplifies large fluctuation contribution to the cross-correlations. In this case, $v_{2j}^{\rho}$ corresponding to Karafuru, 3Landers, and Killer GF collections that fall in the range $v_{2j}^{\rho} \in [0.38, 0.4]$ and $v_{2j}^{\rho}$ corresponding to Goblintown that is equal to $v_{2j}^{\rho} \approx 0.25$ show significantly different values than the average $|v_{2j}^{\rho}| < 0.2$ observed for the other collections. These less-liquid collections experienced a significant drop in capitalization on January 6, 2023, which was also accompanied by an increased number of transactions (inset in Fig~\ref{Cap_N}(b)). The strongest opposite-sign contributions to ${\bf v}_2^{\rho}$ are made by Lil Pudgys and Sappy Seals ($v_{2j}^{\rho} \in [-0.15,-0.2]$).

For the number of transactions time series $N_{\Delta t}$ the results for the expansion coefficients (not shown) differ in that here, for ${\bf C}$ their largest values correspond to the most liquid collections: Mutant Ape Yacht Club, Pudgy Penguins, Bored Ape Yacht Club, Moonbirds, Clone X, Azuki, 0N1 Force, and Milady Maker ($v_{2j} \in [0.15,0.25]$). This is confirmed by the large expansion coefficients of the eigenvector ${\bf v}_2'$ calculated for the filtered matrix ${\bf C}'$, which - despite a slightly different order - form a similar liquid-collection cluster. The largest opposite-sign coefficients $v_{2j} \in [-0.15,-0.2]$ are associated with the least liquid collections: CryptoBatz, Webbland, Coolman's Universe, Smilesss, Project PXN, and 3Landers, while the largest opposite-sign coefficients $v_{2j}' \in [-0.16,-0.22]$ are associated with 3Landers, Alien Frens, Project PXN, Coolman's Universe, Killer GF, CryptoBatz, and Smilesss. In contrast, the expansion coefficients for ${\bf C}^{\rho}(q=4,s=14)$ are dominated by less-liquid collections: Alien Frens, Killer GF, 3Landers, and Coolman's Universe ($v_{2j}^{\rho} \in [0.2,0.31]$), which experienced significant increase in the number of transactions on January 6, 2023. The largest opposite-sign contributions to ${\bf v_2}^{\rho}$ correspond to the following collections: CyberKongz, Hashmasks, and Mutant Ape Yacht Club ($v_{2j}^{\rho} \in [-0.3,-0.2]$).

\subsection{Minimal spanning trees (MST)}

A correlation matrix can be used to form a weighted network where nodes represent collections and edges represent correlations~\cite{tumminello2010correlation}. However, if the number of nodes is significant, such complete networks are difficult to comprehend when plotted and, for the sake of graphical clarity, another useful network representation is recommended, which is minimal spanning tree (MST). It is a subnetwork consisting of $I$ nodes and $I-1$ edges and its key property is that it minimizes the sum of edge weights. In order to create MST, the correlation matrices {\bf C} and ${\bf C}^{\rho}(q,s)$ must be transformed into distance matrices ${\bf D}$ and ${\bf D}^{\rho}(q,s)$, respectively, where the entries, calculated according to the following formula:
\begin{equation}
d_{ij}=\sqrt{2(1-P_{ij}}), \quad d_{ij}^{\rho}(q,s)=\sqrt{2(1-\rho_{ij}(q,s))},
\label{eq::metric-distance}
\end{equation}
are metric distances. The edges minimizing their weight sum are then selected according to Kruskal's or Prim's algorithm~\cite{KruskalJB-1956a,PrimRC-1957a}. From the present work's perspective the structure of an MST can offer a useful insight into the cross-correlation patterns within a market. For instance, a centralized market would be characterized by a star-shaped tree, whereas a market with idiosyncratic dynamics of assets would be characterized by a decentralized structure with extended branches and no significant hub.


\begin{figure}
\includegraphics[width=0.49\textwidth]{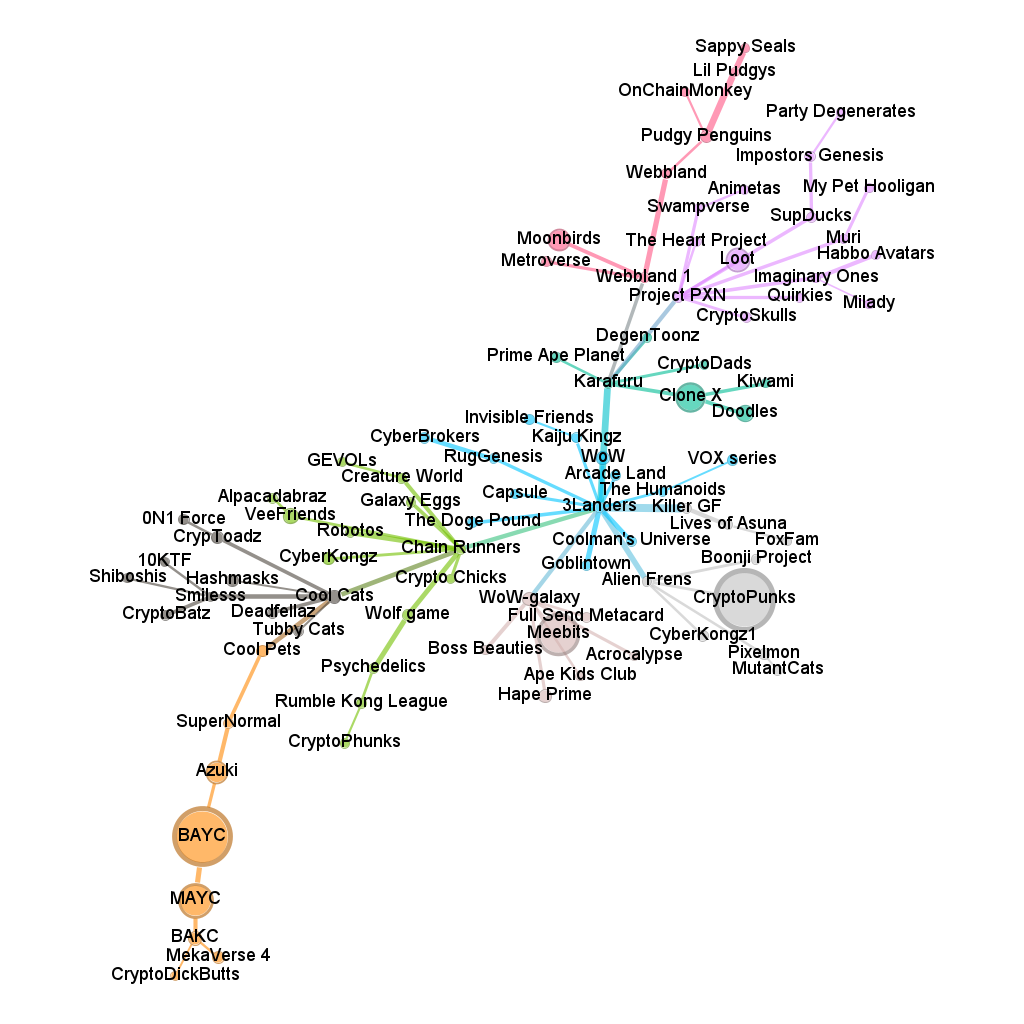}
\caption{MST based on the Pearson correlation matrix ${\bf C}$ for the capitalization increments $c_{\Delta t}$ of different collections. Symbols representing nodes have their size reflecting the collection capitalization on the last day of the period of interest, while node colors denote the communities identified by means of the Louvain method~\cite{Blondel_2008}. Edge thickness shows correlation strength.}
\label{MST_Pc}
\end{figure}


\begin{figure}
\includegraphics[width=0.49\textwidth]{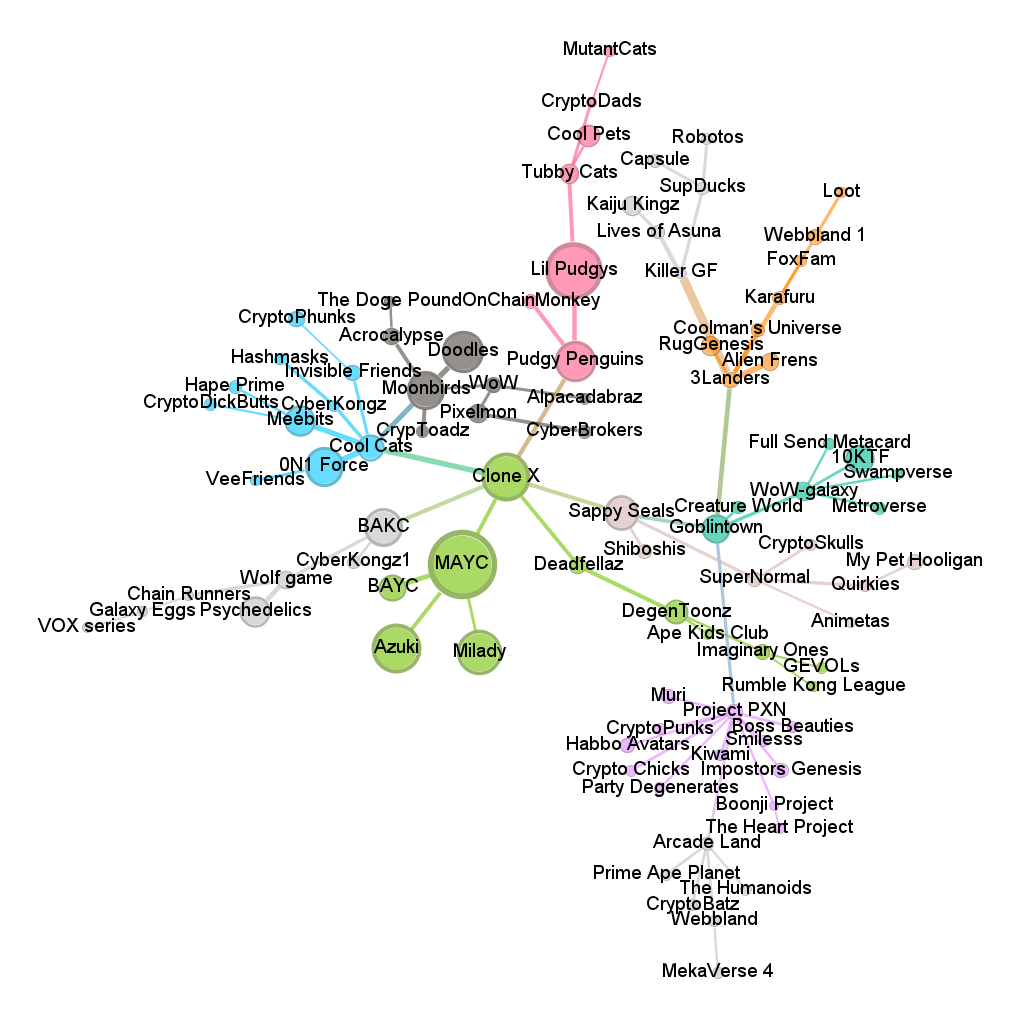}
\caption{Similar MST as in Fig.~\ref{MST_Pc} but for the number of transactions time series $N_{\Delta t}$ and node size represents the total number of transactions in the considered period.}
\label{MST_PN}
\end{figure}


\begin{figure}
\includegraphics[width=0.49\textwidth]{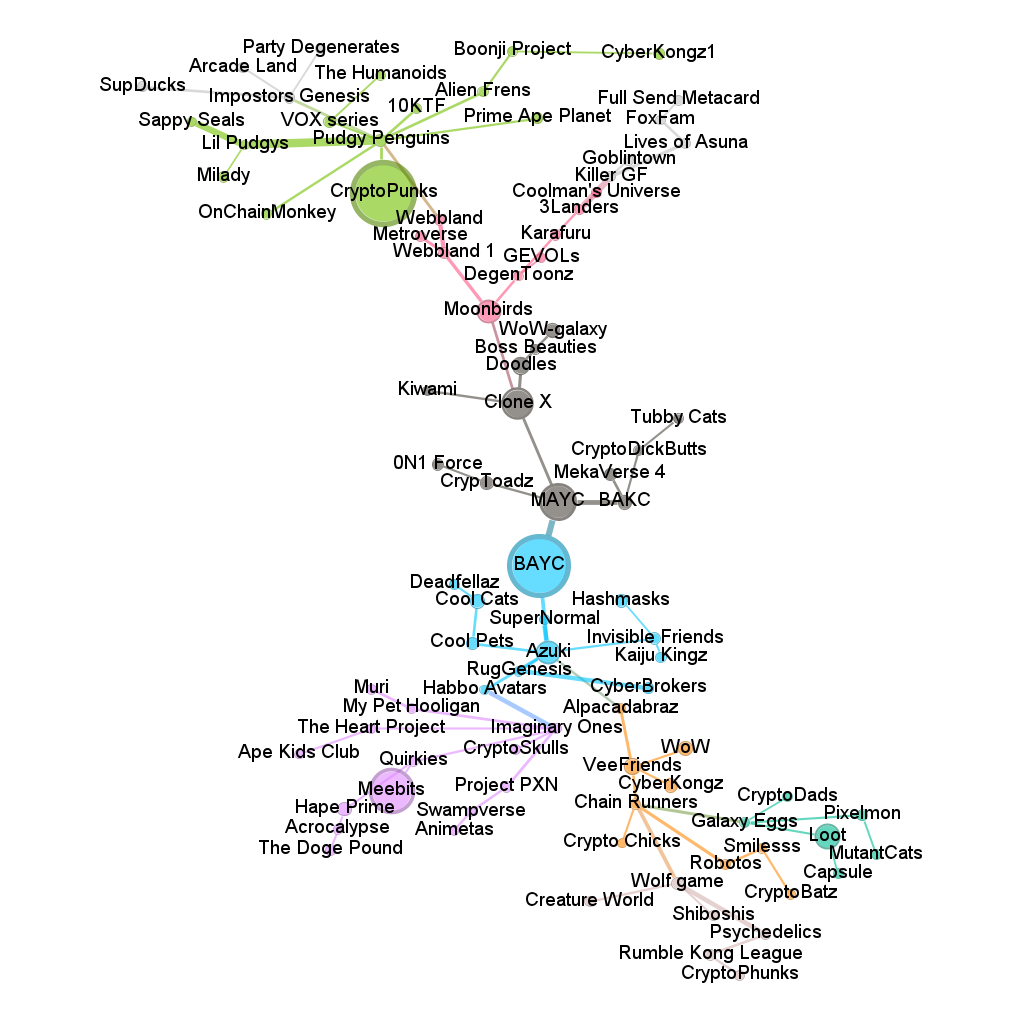}
\caption{MST for the same $c_{\Delta t}$ dataset as in Fig.~\ref{MST_Pc} but for the filtered correlation matrix ${\bf C}'$ for the original time series.}
\label{MST_PZc}
\end{figure}


\begin{figure}
\includegraphics[width=0.49\textwidth]{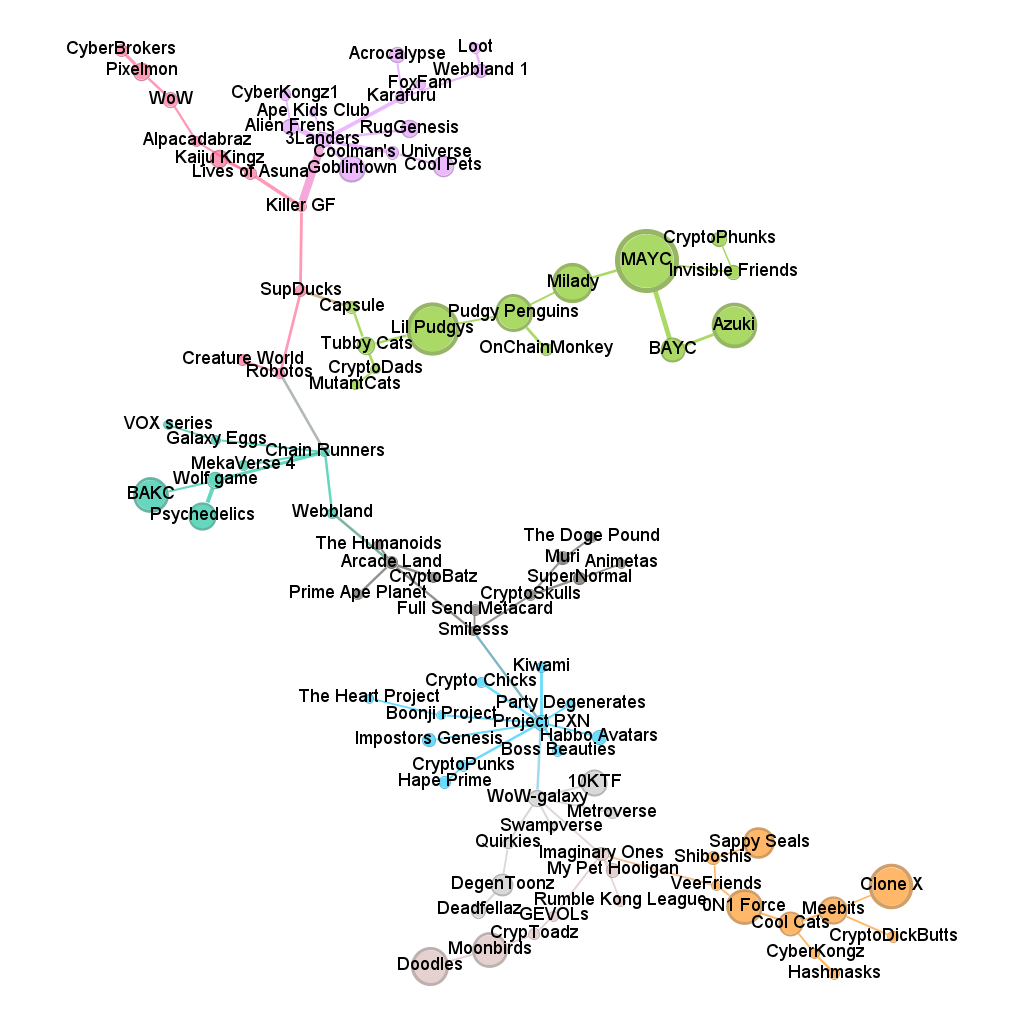}
\caption{MST for the same $N_{\Delta t}$ dataset as in Fig.~\ref{MST_PN} but for the filtered correlation matrix ${\bf C}'$ for the original time series.}
\label{MST_PZN}
\end{figure}


\begin{figure}
\includegraphics[width=0.49\textwidth]{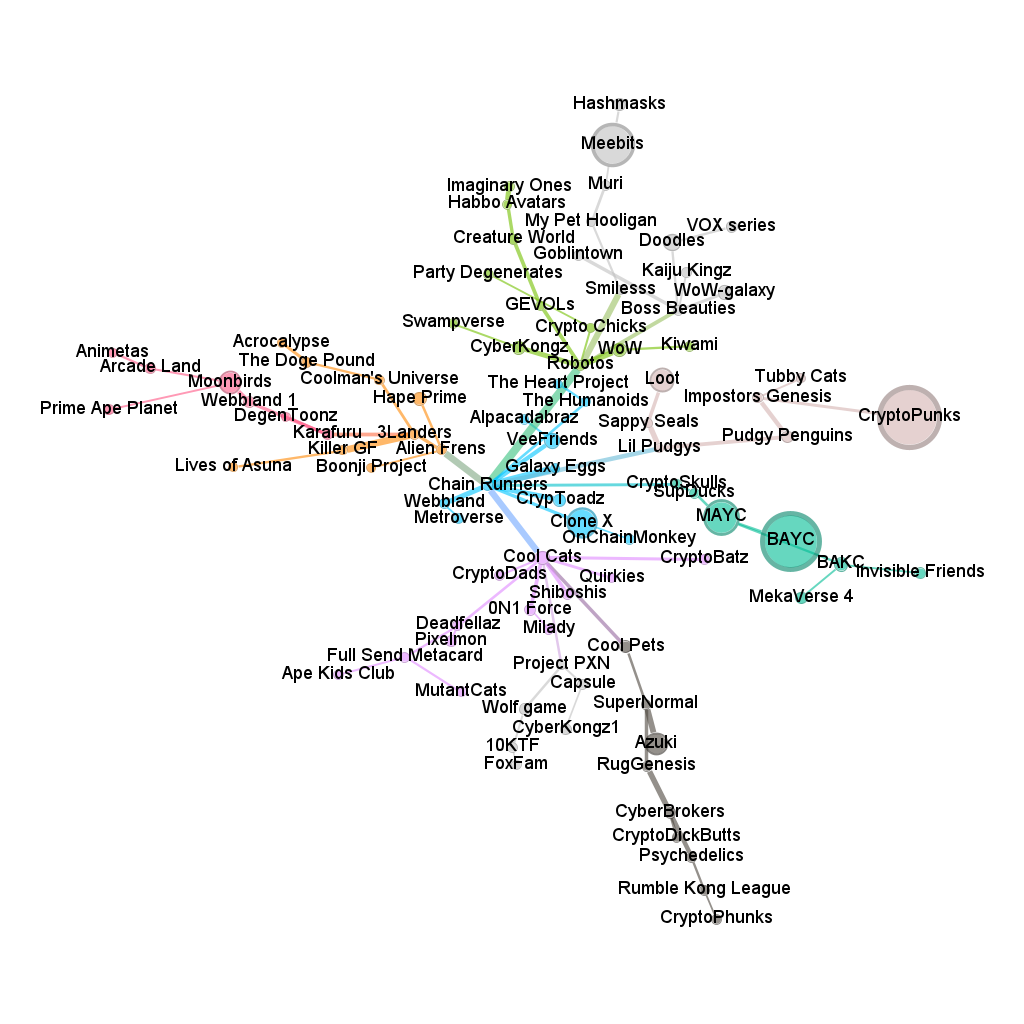}
\caption{MST based on the $q$-dependent detrended correlation matrix ${\bf C}^{\rho}(q=4,s=14)$ for the capitalization increments $c_{\Delta t}$.}
\label{MST_q4s14c}
\end{figure}


\begin{figure}
\includegraphics[width=0.49\textwidth]{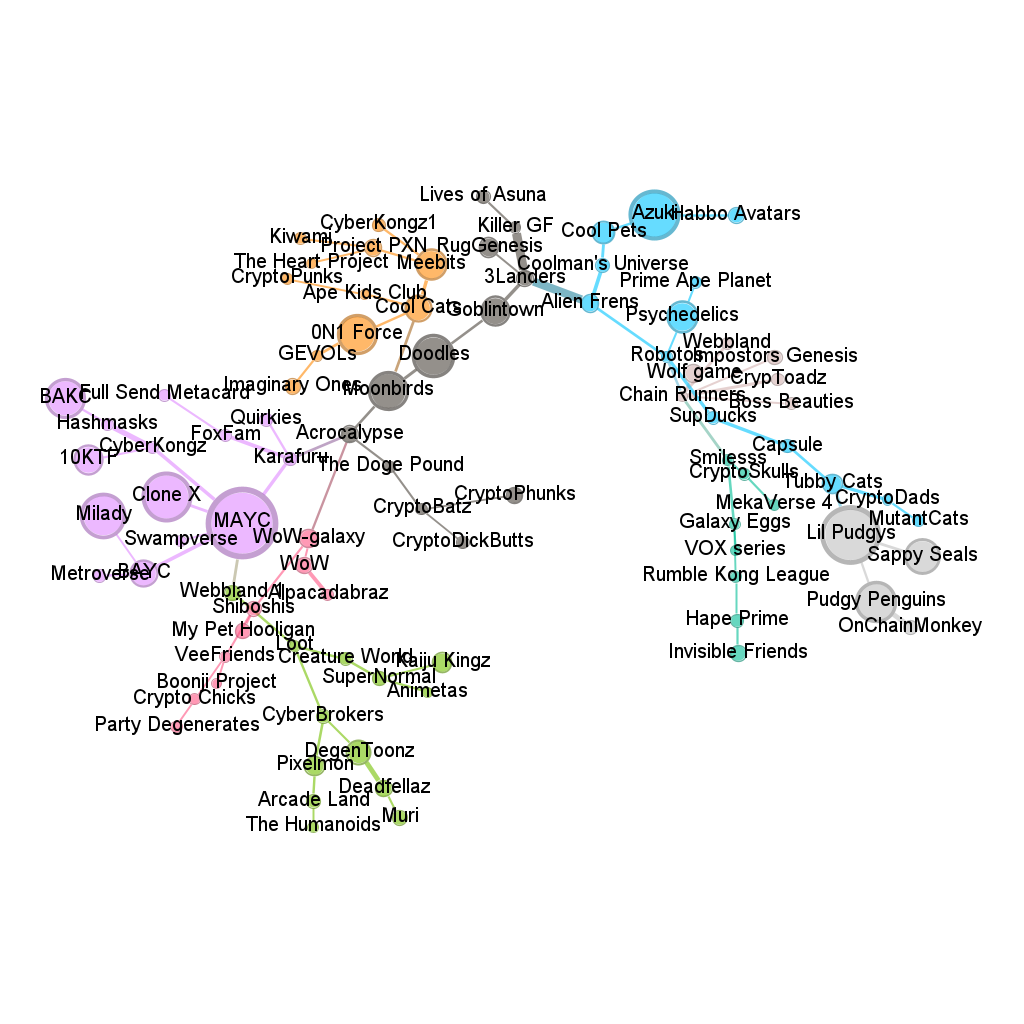}
\caption{MST based on the $q$-dependent detrended correlation matrix ${\bf C}^{\rho}(q=4,s=14)$ for the number of transactions $N_{\Delta t}$.}
\label{MST_q4s14N}
\end{figure}

The MST presented in Fig.~\ref{MST_Pc} was constructed from the correlation matrix ${\bf C}$ based on the Pearson coefficients between the time series of the collection capitalization increments $c_{\Delta t}$. The MST is characterized by a rather distributed structure with no dominant hub. This remains in agreement with the observation coming out of Fig.~\ref{V1_Cap} that the market factor represented by ${\bf v}_1$ does not comprise all the collections with the same strength. The strongest local center is the node corresponding to 3Landers with a degree of $\delta=14$, which is also the central node for a community of 15 collections. Among other minor centers one can point out to Project PXN, Karafuru, Chain Runners, and Cool Cats. This is remarkably different than in the case of other financial markets like the stock market, Forex, or cryptocurrency market where the structure is predominantly centralized with a well-distinguished hub~\cite{watorek2021,entropy2021b,WatorekM-2023b}. Such a structure is not restricted to capitalization data, but also occurs in the case of the number of transactions $N_{\Delta t}$ -- see Fig.~\ref{MST_PN}. Here the most connected node, which is Project PXN, has degree of $\delta=12$ and there exist a few other small centers like 3Landers, Cool Cats and Clone X.

It is interesting to look at the MST structure after removing the variance component corresponding to $\lambda_1$. For both $c_{\Delta t}$ (Fig.~\ref{MST_PZc}) and $N_{\Delta t}$ (Fig.~\ref{MST_PZN}), MST remains decentralized as one might expect from the structure of the corresponding eigenvector ${\bf v}_2'$ in Fig.~\ref{V2_Cap}. The most connected nodes are Pudgy Penguins for $c_{\Delta t}$ and Project PXN for $N_{\Delta t}$, both with $\delta=10$. The secondary cluster centers are Azuki for $c_{\Delta t}$ and 3Landers for $N_{\Delta t}$, both with $\delta=7$. 

The $c_{\Delta t}$-based MST created from ${\bf C}^{\rho}(q=4,s=14)$ focuses on the correlation structure among large capitalization increments. Consistently with the fat-tailed distribution of the matrix elements depicted in Fig.~\ref{fig::cc}(b), where there are few large elements and many small ones, the MST reveals a distributed structure with a few small clusters centered at Chain Runners ($\delta=11$), Robotos ($\delta=9$) and Cool Cats ($\delta=7$) -- see Fig.~\ref{MST_q4s14c}. An evident property of the MSTs for $c_{\Delta t}$ is that the nodes corresponding to the collections with the largest capitalization (large circles), with the exception of the Ape collections, do not cluster together. Especially the collection with the largest capitalization -- CryptoPunks -- is peripheral in all the cases. This effect can be viewed as a contrasting one in respect to the other financial markets, where the largest stocks, currencies, and cryptocurrencies were identified as central nodes~\cite{KwapienJ-2012a,watorek2021,entropy2021b,WatorekM-2023b}. For $N_{\Delta t}$, the corresponding MST is even more dispersed with the most connected nodes being Chain Runners ($\delta=7$) and Mutant Ape Yacht Club ($\delta=6$) -- see Fig.~\ref{MST_q4s14N}. The cluster centered around the latter contains the collections with the largest number of transactions.

\begin{figure}
\includegraphics[width=0.49\textwidth]{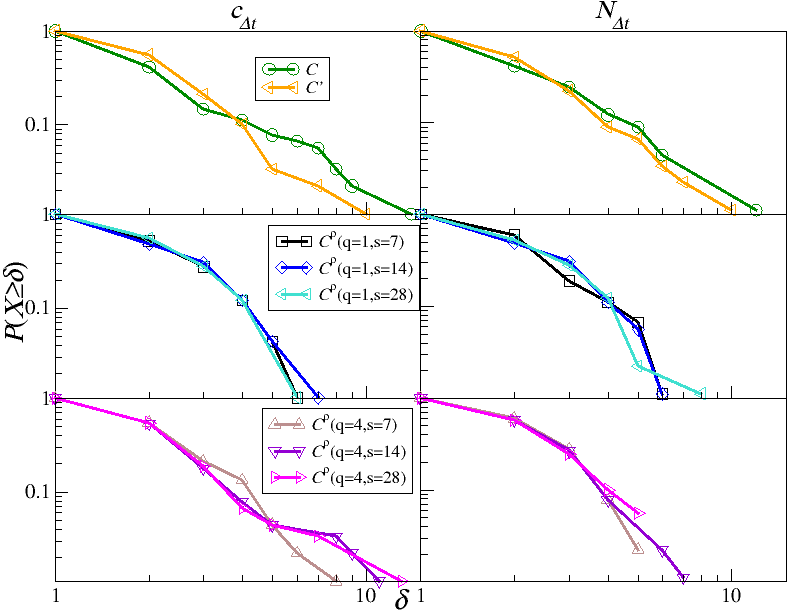}
\caption{Complementary cumulative distribution function $P(X \geq \delta)$ of the node degrees $\delta$ of MSTs created from the collection capitalization increments $c_{\Delta t}$ (left) and the number of transactions $N_{\Delta t}$ (right). The correlation matrices based on the Pearson coefficients ${\bf C}$ (top) and the $q$-dependent detrended correlation coefficients ${\bf C}^{\rho}(q=1,s)$ (middle) and ${\bf C}^{\rho}(q=4,s)$ (bottom) were used.}
\label{fig::nodedegreePDF}
\end{figure}

The decentralized structure of the MSTs presented in Figs.~\ref{MST_Pc}-\ref{MST_q4s14N} can also be inferred from the node degree cumulative distributions shown in Fig.~\ref{fig::nodedegreePDF} where no node stands out as an outlier. In the case of the MSTs based on ${\bf C}$ (top panels), even a trace of a power-law dependence can be seen (straight lines in the double logarithmic scale plots). Such distributions of $\delta$ define the scale-free networks~\cite{Barabasi2003} and were also reported for foreign currency exchange rates~\cite{gorski2008scale} and cryptocurrencies~\cite{POLOVNIKOV2020,watorek2021}. No similar result can be seen for ${\bf C}^{\rho}(q=1,s)$ (middle panels) and ${\bf C}^{\rho}(q=4,s)$ (bottom panels).


\section{Summary}

Despite its young age and unique trading dynamics, the NFT market exhibits many similarities with the cryptocurrency market and the traditional financial markets like the stock market and Forex. Among such similarities one can point out to long range memory and fat tailed probability distribution functions of some observables (e.g., absolute values of logarithms of the capitalization increments and the number of transactions in time unit). However, there are differences as well. The collection cross-correlations tend to be lower than their counterparts in the other markets, which manifests itself in the magnitude of the largest eigenvalue $\lambda_1$ of the corresponding correlation matrices (both those based on the Pearson correlations and the $q$-dependent detrended ones), yet they markedly deviate from random ones. Another difference is the fact that only a fraction of the total number of the collections contribute to a market factor, while the remaining ones show more independent behavior and can be grouped in clusters. Moreover, there is a significant negative contribution of some collections to the overall dynamics, which is quantified in the negative vector expansion coefficients in the eigenvector related to $\lambda_1$. Such anti-correlation was not so frequently observed in the case of cryptocurrencies, for example. Another interesting atypical property of the NFT market is strong cross-correlations between large fluctuations of the studied observables among less liquid collections, especially in the case of the number of transactions in time unit.

An additional insight into the cross-correlation structure of the NFT market has been given by a network approach that utilized minimal spanning trees. They reveal a largely decentralized structure of the respective MSTs that lack a dominant node, which remains in a striking contrast with the cryptocurrency market not to mention the traditional markets. Despite the fact that there exist large discrepancies in capitalization of the collections and the number of transactions, the price evolution of the largest, most frequently traded collections are not necessarily associated with the respective characteristics of the less priced collections. This is yet another disparity between the NFT collections and the other financial assets, where the highly capitalized assets dominate the evolution of a market and tend to be coupled to each other.

These differences go far beyond a lesser maturity and the associated lower liquidity of the NFT assets and they could stem from the unique trading mechanisms that are distinct from those in the other markets. For example, there is no order book here and, what is even more important, token non-fungibility makes performing simple arbitrage impossible.  Consequently, the information transmission mechanism is slower both within the same collection and between different collections. All these factors lead to the observed weaker cross-correlations among the tokens than in the case of cryptocurrencies and traditional financial assets.

\begin{acknowledgments}
This research was funded in whole or in part by National Science Centre, Poland 2023/07/X/ST6/01569.
\end{acknowledgments}

\section*{Data Availability Statement}

The data is freely available from CryptoSlam!~\cite{CryptoSlam} portal.

\section{Appendixes}

\begin{table*}[]
\caption{NFT collections considered in this study. $K$ - collection capitalization on the last day from the dataset, $N_{\rm tot}$ - total number of transactions, $S$ - collection size measured by the number of circulated tokens, and $\%0_{\Delta t=1h}$ - the fraction of hours without transactions.}
\begin{tabular}{|l|l|l|l|l|l|l|l|l|l|}
\hline
\textbf{Name}               & $K [10^6$\$] & $N_{\rm tot}$ & $S$  & $\%0_{\Delta t=1h}$ & \textbf{Name}                  & $K [10^6$\$]  & $N_{\rm tot}$ & $S$  & $\%0_{\Delta t=1h}$ \\ \hline
0N1 Force                   & 51.3         & 28565      & 7766          & 54\%           & Invisible Friends              & 47.3         & 7432       & 4931            & 67\%          \\ \hline
10KTF                       & 21.8         & 18901      & 25826           & 46\%            & Kaiju Kingz                    & 35.1         & 10986      & 7289            & 55\%            \\ \hline
3Landers                    & 29.4         & 7498       & 8650            & 69\%            & Karafuru                       & 34.1         & 4801       & 5415            & 77\%            \\ \hline
Acrocalypse                 & 9.4          & 8302       & 10010            & 74\%           & Killer GF                      & 20.3         & 2944       & 6618            & 85\%           \\ \hline
Alien Frens                 & 28.3         & 8778       & 9569            & 58\%           & Kiwami                         & 18.8         & 3313       & 9392            & 84\%           \\ \hline
Alpacadabraz                & 17.0         & 2965       & 9287            & 87\%           & Lil Pudgys                      & 10.5          & 44984      & 17853            & 28\%           \\ \hline
Animetas                    & 16.1         & 1911       & 9672           & 91\%           & Lives of Asuna                 & 18.6         & 5117       & 9771            & 80\%           \\ \hline
Ape Kids Club               & 19.8         & 1688       & 9466            & 92\%           & Loot                           & 323.8        & 2553       & 6153            & 91\%           \\ \hline
Arcade Land                 & 38.2         & 5040       & 8677            & 76\%           & Meebits                        & 724.1        & 20741      & 13345            & 53\%           \\ \hline
Azuki                       & 292.1        & 37565      & 9718            & 36\%           & MekaVerse 4                    & 77.3      & 3151       & 8576            & 85\%           \\ \hline
Boonji Project              & 20.0         & 1588       & 11004            & 91\%           & Metroverse City Block          & 29.1         & 3969       & 8930            & 85\%           \\ \hline
Bored Ape Kennel Club (BAKC)      & 105.5        & 27892      & 7227            & 43\%           & Milady                         & 31.1         & 33123      & 8466            & 41\%           \\ \hline
Bored Ape Yacht Club (BAYC)        & 1082.7       & 17383      & 9290            & 48\%           & Moonbirds                      & 291.3        & 27797      & 8747            & 48\%           \\ \hline
Boss Beauties               & 21.3         & 3117       & 8858            & 83\%           & Muri                           & 21.5         & 6188       & 9926            & 76\%           \\ \hline
Capsule                     & 28.9         & 4899       & 9668           & 85\%           & Mutant Ape Yacht Club (MAYC)          & 559.5        & 55694      & 15809            & 18\%           \\ \hline
Chain Runners               & 28.3         & 2204       & 9915            & 91\%           & MutantCats                     & 22.5         & 3340       & 9940            & 89\%           \\ \hline
Clone X                     & 445.8        & 36962      & 12076            & 31\%           & My Pet Hooligan                & 20.3         & 6617       & 7779            & 68\%           \\ \hline
Cool Cats                   & 106.8        & 16948      & 8840           & 54\%           & OnChainMonkey                  & 25.6         & 5422       & 8244            & 75\%           \\ \hline
Cool Pets                   & 75.2         & 13299      & 14377            & 78\%           & Party Degenerates              & 22.0         & 3035       & 9705            & 88\%           \\ \hline
Coolman's Universe          & 24.8         & 5269       & 9130            & 55\%           & Pixelmon                       & 47.5         & 10499      & 9549            & 70\%           \\ \hline
Creature World              & 37.3         & 4298       & 9814           & 80\%           & Prime Ape Planet               & 45.7         & 3183       & 7598            & 86\%           \\ \hline
Crypto Chicks                & 14.6         & 2863       & 9944            & 78\%           & Project PXN                    & 32.9         & 7878       & 9453            & 66\%           \\ \hline
CrypToadz                   & 84.0         & 4107       & 6452            & 83\%           & Psychedelics Anonymous Genesis & 25.5         & 20876      & 9474            & 47\%           \\ \hline
CryptoBatz by Ozzy Osbourne & 28.2         & 3158       & 8734            & 86\%           & Pudgy Penguins                 & 70.9         & 29869      & 8897            & 41\%           \\ \hline
CryptoDads                  & 17.4         & 2168       & 9872            & 89\%           & Quirkies                       & 16.9         & 4529       & 4735            & 79\%           \\ \hline
CryptoDickButts S3          & 14.6         & 4108       & 5140            & 81\%           & Robotos                        & 19.1         & 2624       & 9885            & 87\%           \\ \hline
CryptoPhunks V2             & 27.4         & 7813       & 7450           & 73\%           & RugGenesis                     & 19.6         & 10554      & 17200            & 67\%           \\ \hline
CryptoPunks                 & 1147.4       & 2964       & 6960            & 83\%           & Rumble Kong League             & 34.1         & 2612       & 9897            & 87\%           \\ \hline
CryptoSkulls                & 24.4         & 3283       & 7680            & 81\%           & Sappy Seals                    & 13.9         & 24449      & 9737            & 48\%           \\ \hline
CyberBrokers                & 53.4         & 4930       & 7744            & 79\%           & Shiboshis                      & 26.1         & 5767       & 6882            & 73\%           \\ \hline
CyberKongz VX 1            & 51.2         & 4734       & 11952            & 82\%           & Smilesss                       & 15.7         & 2565       & 7296            & 87\%           \\ \hline
CyberKongz                  & 100.3        & 3101       & 3304            & 77\%           & SupDucks                       & 27.4         & 4131       & 9950            & 79\%           \\ \hline
Deadfellaz                  & 33.0         & 8139       & 9197            & 61\%           & SuperNormalbyZipcy             & 24.7         & 6057       & 7222            & 79\%           \\ \hline
DegenToonz                  & 14.2         & 14989      & 8210            & 56\%           & Swampverse                     & 11         & 2321       & 9523            & 91\%           \\ \hline
Doodles                     & 174.0        & 30860      & 8690            & 45\%           & The Heart Project              & 15.1         & 2003       & 9651            & 83\%           \\ \hline
FoxFam                      & 15.3         & 3189       & 9836            & 87\%           & The Humanoids                  & 18.4         & 2406       & 9802            & 91\%           \\ \hline
Full Send Metacard          & 27.6         & 3452       & 9315            & 82\%           & The Doge Pound                 & 39.6         & 3030       & 9919            & 87\%           \\ \hline
Galaxy Eggs                 & 20.4         & 2547       & 9135            & 89\%           & Tubby Cats                     & 22.7         & 10509      & 19468            & 70\%           \\ \hline
GEVOLs                      & 14.5         & 2817       & 8478            & 85\%           & VeeFriends                     & 149.8        & 2464       & 9106            & 85\%           \\ \hline
Goblintown                  & 28.3         & 19183      & 8986            & 57\%           & VOX Series 1                   & 53.3          & 2050       & 8702            & 87\%           \\ \hline
Habbo Avatars               & 12.3         & 6836       & 10610           & 67\%           & Webbland 1                     & 39.2         & 6975       & 9058            & 85\%           \\ \hline
Hape Prime                  & 100.2        & 4550       & 7068            & 77\%           & Webbland                       & 34.9         & 3346       & 8498           & 79\%           \\ \hline
Hashmasks                   & 84.5         & 3527       & 16336            & 84\%           & Wolf Game                      & 55.7         & 9233       & 11518            & 70\%           \\ \hline
Imaginary Ones              & 21.0         & 6805       & 8456            & 71\%           & World of Women Galaxy (WoW-galaxy)          & 115.4        & 9542       & 14360            & 71\%           \\ \hline
Impostors Genesis           & 37.5         & 6222       & 9656            & 69\%           & World of Women (WoW)                & 102.2        & 7415       & 9397            & 58\%           \\ \hline
\end{tabular}
\label{NFTnames}
\end{table*}

\nocite{*}
\bibliography{ref}

\end{document}